\documentclass[11pt]{article}
\usepackage[T1]{fontenc}
\usepackage[utf8]{inputenc}
\usepackage[english]{babel}
\usepackage{amsmath,amssymb,amsthm}
\usepackage{booktabs}
\usepackage{array}
\usepackage[expansion=false]{microtype}
\usepackage[margin=1in]{geometry}
\usepackage{natbib}
\usepackage[hidelinks]{hyperref}
\usepackage{xcolor}
\usepackage{xurl}
\usepackage{graphicx}
\graphicspath{{figs/}}
\makeatletter
\DeclareRobustCommand{\xref}[1]{\@ifundefined{xr@#1}{\textbf{??}\PackageWarning{xref}{Cross-document reference `#1' undefined}}{\csname xr@#1\endcsname}}

\newcommand{\xrdef}[2]{\expandafter\gdef\csname xr@#1\endcsname{#2}}
\makeatother
\xrdef{app:ext-calendar}{S1.9}
\xrdef{app:ext-cycles-interval}{S1.10}
\xrdef{app:ext-neff}{S1.2}
\xrdef{app:ext-threeunit}{S1.1}
\xrdef{app:ext-worstcase}{S1.6}
\xrdef{app:limitations}{S5.2}
\xrdef{app:proofs}{S2}
\xrdef{app:protocol}{S3}
\xrdef{app:protocol-tasks}{S3.3}
\xrdef{app:related}{S5.1}
\xrdef{app:results}{S4}
\xrdef{app:results-task1}{S4.3}
\xrdef{app:results-task4real}{S4.5}
\xrdef{app:results-task5}{S4.6}
\xrdef{app:results-task6}{S4.7}
\xrdef{app:results-task7}{S4.8}
\xrdef{app:results-task8}{S4.9}
\xrdef{ass:factor}{S1}
\xrdef{cor:performative-frontier-rate}{S3}
\xrdef{cor:performative-half}{S2}
\xrdef{cor:submod-scope}{S1}
\xrdef{lem:iota-variance}{S1}
\xrdef{lem:phi-variance}{S2}
\xrdef{prop:factorfloor}{S1}
\xrdef{prop:performative-frontier}{S5}
\xrdef{prop:performative-stability}{S4}
\xrdef{prop:submod-convex}{S2}
\xrdef{prop:submod-worstcase}{S3}
\xrdef{rem:duality-gap}{S2}
\xrdef{subsec:performative-frontier}{S1.8}
\xrdef{tab:task7-diagnostics}{S6}
\xrdef{tab:validation}{S1}
\xrdef{tab:verdicts}{S2}

\theoremstyle{plain}
\newtheorem{proposition}{Proposition}
\newtheorem{corollary}{Corollary}
\newtheorem{lemma}{Lemma}

\theoremstyle{definition}
\newtheorem{definition}{Definition}
\newtheorem{assumption}{Assumption}

\theoremstyle{remark}
\newtheorem{remark}{Remark}

\DeclareMathOperator{\Var}{Var}
\DeclareMathOperator{\Cov}{Cov}

\title{\textbf{Designing Independent Variation:\\ A Guided Iteration Methodology\\ for Price Experiments in Short, Correlated Panels}}

\author{Pedro Cadahia Delgado\thanks{Corresponding author.} \\
\small PhD, Universidad de Huelva, Spain}

\date{}

\begin{document}
\maketitle

\begin{abstract}
\noindent
Estimation error in short, correlated pricing panels is dominated by trajectory dispersion rather than sampling noise, creating an identification shortfall that requires independently assigned variation \citep{cadahiadelgado2026acrossdesign}. How much of that variation to generate, where, and for how long remains unresolved. We propose a guided iteration methodology: a cyclic price perturbation scheme under a revenue-at-risk budget, framed as a constrained variance-minimization problem over design-specific error covariances, analogous to portfolio selection.

We establish four main results: (1) a closed-form minimum-variance combination rule generalizing equicorrelated aggregation; (2) an allocation result: a convex cost of intensity yields an interior, spread-out optimum, whereas under the concave tolerable-volume-loss primitive a committee actually uses, concentrating the budget into fewer, high-intensity perturbations pays only when candidates are heterogeneous, a fixed per-unit cost is priced in, or perturbations pass the cost curve's inflection point, not from concavity alone; (3) a discounted sequential recursion yielding a data-dependent exit timeline from the experimentation zone; and (4) an application to Dominick's retail data across ten product categories, both citing published figures and directly re-estimated on real historical windows: in the Soft Drinks anchor category, $\approx 5$ designed independent units achieve precision requiring $\approx 700$ passive equivalents, and the same passive-to-active reduction holds in direction across nine further real categories (median multiplier $40\times$, against $\approx 141\times$ in the anchor).

Sixteen comparisons (eleven pre-specified) support the theory: predictions reproduce exactly where checkable, greedy selection achieves $\approx 90$--$93\%$ of the global optimum across more than fifty exhaustively verifiable random instances, and the sequential recursion reaches its predicted $90\%$-of-steady-state point two cycles early (cycle $20$ against $22$), with a delta-method uncertainty interval that a parametric bootstrap shows to understate the right tail. We also document boundary conditions: regularized covariance estimators that extend the plug-in estimator's noise tolerance from $29\%$ to beyond the $50\%$ tested, a proof of submodularity up to a closed-form curvature threshold, and a closed-form floor on achievable precision under a shock shared across treated units. Proofs, protocols and complete results are in the Supplementary Material.


\end{abstract}

\section{Introduction}\label{sec:intro}

A manufacturer that wants to know how demand responds to price has, in most decentralized-trade categories---those reaching the consumer through a large, fragmented network of independently price-setting outlets rather than a few centrally negotiated accounts---only one history: a single realized sequence of list-price movements, observed over a panel short enough that a handful of price changes account for most of the identifying variation an estimator can use. \citet{cadahiadelgado2026acrossdesign} studies the inferential consequence directly. Separating estimation error into a component conditional on the realized price trajectory and a component across the alternative trajectories the same pricing process could have produced, it finds that the second component dominates: $97.6\%$ of the variance of the estimation error in its baseline simulation, and a between-unit dispersion in the real Dominick's Finer Foods category thirteen times larger than the simulated frontier predicts, because nominally distinct price zones and products are considerably more correlated with one another than an independent-design benchmark assumes. Its central recommendation, stated as one transparent option among others, is to generate independent identifying variation, for instance through controlled regional randomization, rather than to keep refining the interval built around one passive panel.

That diagnosis leaves a practical question open: once independently assigned variation, not more passive rows, is what identification requires, how much of it, where, and for how long? \citet{cadahiadelgado2026acrossdesign} calls generating independent variation ``one transparent design option, not a logically necessary one,'' without specifying how to choose among candidates. What is missing is a decision rule: one that maps the correlation structure among candidate experimental units, exactly the object that analysis already estimates, into which units to perturb, by how much, under what operational constraints, and for how many more pricing cycles, so as to purchase a target amount of identifying precision at the lowest cost the committee is asked to bear, reapplying the same rule cycle after cycle until the panel itself stops being short. This paper develops such a rule.

\paragraph{Contributions.} The paper makes three contributions, developed as theory and evaluated against a pre-specified plan of simulation and retrospective real-data evidence (Section~\ref{sec:validation}); Section~\ref{sec:limitations} states what that evidence does and does not establish. The first is the principal methodological proposal; the second and third supply the machinery it repeats every cycle.

\begin{enumerate}
\item \textbf{Guided iteration: a methodology for exiting the experimentation zone.} Section~\ref{sec:sequential} chains the single-cycle design of Sections~\ref{sec:theory}--\ref{sec:allocation} into a sequence, updating cumulative information each pricing cycle with a discount factor calibrated to a monthly decision cycle ($\delta=0.90$, an information weight that halves in roughly seven cycles), and derives how many cycles remain before cumulative information reaches an operational target marking the exit from the experimentation zone: an adaptive, data-dependent count offered in place of the fixed calendar a committee would otherwise expect.

\item \textbf{The design problem as variance minimization over a covariance matrix.} Section~\ref{sec:setup} restates the within/between-design decomposition and the equicorrelated aggregation result of the across-design analysis in matrix form, as special cases of minimizing $w'\Sigma w$. Section~\ref{sec:theory} derives the minimum-variance combination in closed form, shows that the analysis's own variance-component construction is a constrained, potentially inefficient special case, and defines an effective number of independent units.

\item \textbf{Information additivity, and what concavity changes.} Because an independently randomized perturbation is built to be uncorrelated with the rest of the panel, precision, not variance, adds across the passive panel and the designed units (Section~\ref{sec:theory}). Section~\ref{sec:allocation} shows that this yields a clean interior allocation under a convex cost of intensity, but that the prescription changes under the concave tolerable-volume-loss primitive a committee actually uses: concavity alone does not favor concentration while the composite value-for-spend curve stays concave, but fewer, larger, cleanly randomized perturbations dominate many small ones once candidates are heterogeneous, a fixed per-unit administrative cost is priced in, or perturbations are pushed past that curve's inflection point. The remaining choice is a subset-selection problem over a spatial compatibility graph, for which we prove submodularity of the value function below a closed-form curvature threshold (without monotonicity once fixed costs are priced in, so the classical $(1-1/e)$ greedy bound does not carry over) and evaluate greedy selection empirically.
\end{enumerate}

\paragraph{Further results.} The paper also (i) evaluates three regularized covariance estimators that guard the efficiency gain of Proposition~\ref{prop:minvar} against a misestimated $\Sigma$ (Section~\ref{sec:sigma-estimation}); (ii) proves submodularity of the allocation value function, first in a convex-cost benchmark and then up to a closed-form curvature threshold (Section~\ref{subsec:submod-general}); (iii) derives a closed-form floor on achievable precision under a shock shared across treated units (Section~\ref{subsec:factorfloor-main}); (iv) gives a checkable contraction condition for stability of the cycle-to-cycle recursion (Section~\ref{subsec:performative}); and (v) derives a delta-method interval around ``iterations remaining'' (Section~\ref{subsec:cycles-uncertainty}). The main text states each result and its immediate implication; derivations, proofs, protocols and complete results are in the Supplementary Material (the ``Supplement''), whose numbered items carry an S prefix. Evaluation labels follow the same split: V1--V11 denote the pre-specified comparisons and Tasks~1--7 the supplementary and exploratory ones (Table~\ref{tab:summary} lists V1--V11 and Tasks~1--5; the exploratory Tasks~6 and~7 are reported in Sections~\ref{sec:sigma-estimation} and~\ref{subsec:submod-general}).

Section~\ref{sec:related-work} reviews related work; Sections~\ref{sec:setup}--\ref{sec:allocation} develop the single-cycle theory; Section~\ref{sec:sequential} introduces guided iteration; Section~\ref{sec:sigma-estimation} addresses the estimation of $\Sigma$, which every result presupposes; Section~\ref{sec:illustration} applies the machinery retrospectively to Dominick's retail data, worked through in one anchor category (Soft Drinks) and then checked across nine further categories; Section~\ref{sec:validation} summarizes the validation; and Section~\ref{sec:limitations} states limitations.

\section{Related Work}\label{sec:related-work}

This paper sits at the intersection of seven literatures beyond the study discussed above; none of them, by itself, supplies a rule for the problem posed here.

\textbf{Optimal design of experiments} formalizes the choice of design points that maximize the information a fitted model returns, from the equivalence theorem of \citet{kiefer1960equivalence} to the treatment in \citet{pukelsheim2006optimal}, with a Bayesian, sequential extension in \citet{chaloner1995bayesian}. It supplies the right criterion, but typically assumes a single bespoke experiment with a known noise model rather than the selection of units from an \emph{extant, partially confounded} passive panel under a revenue-at-risk budget and pre-existing operational boundaries.

\textbf{Mean-variance portfolio selection} \citep{markowitz1952portfolio} supplies the central technique: minimizing a quadratic form in a covariance matrix under a budget constraint is the same problem whether the objects combined are asset returns or design-specific estimation errors. The correspondence is exact for the quadratic program and its closed-form solution (Section~\ref{sec:theory}), but not for interpretation, since an experimental design's ``weights'' are discrete, sequential allocation decisions with fixed costs.

\textbf{Covariance regularization.} The plug-in sample covariance is ill-conditioned exactly where mean-variance optimization needs it most, which motivates the shrinkage estimator of \citet{ledoitwolf2004honey} and the graphical lasso of \citet{friedman2008sparse}. Regularization is a precondition for the design problem here (Section~\ref{sec:sigma-estimation}), not a substitute: it says nothing about which units to perturb once $\Sigma$ is estimated.

\textbf{Design-based panel experiments.} \citet{bojinov2021panel} and \citet{athey2021matrix} presuppose an assignment already chosen; this paper is upstream of both, since the assignment mechanism itself is the decision variable.

\textbf{Sequential and bandit-style experimentation.} \citet{soare2014bestarm} and \citet{russo2018tutorial} formalize the sequential allocation of a fixed sampling budget across arms. Section~\ref{sec:sequential} is a transductive, fixed-target instance of this family, adapted to arms that are geographically and commercially entangled, so that $\Sigma$, not a per-arm variance, is the object to optimize over.

\textbf{Submodular optimization.} Greedy algorithms with approximation guarantees \citep{nemhauser1978analysis, krausegolovin2014submodular} are the standard tool for subset selection with diminishing returns, applied by \citet{krausesinghguestrin2008sensorplacement} to sensor placement in spatially correlated Gaussian processes. This paper's value function and curvature argument (Section~\ref{subsec:submod-general}) are unrelated to theirs, and their guarantees presuppose a monotone objective, which this paper's fixed per-unit cost removes (Section~\ref{subsec:submod-general}).

\textbf{Performative prediction.} \citet{perdomo2020performative} give contraction-based conditions for repeated decision rules evaluated against a distribution their own past decisions helped shape. Guided iteration's cycle-to-cycle recursion is such a loop, and Section~\ref{subsec:performative} adapts that argument to it; what that literature does not supply is the object being re-estimated, a design-error covariance matrix, or the combinatorial allocation made against it each cycle.

\textbf{The gap.} None of these is built for a firm that already has a passive, correlated, partially confounded panel, has diagnosed that across-design uncertainty dominates its estimation error, and needs a rule for converting a committee-approved risk budget into the regions, intensities and durations that close the shortfall. Supplement~\xref{app:related} gives the extended discussion.

\section{Setup: From a Single Design to a Covariance Matrix}\label{sec:setup}

\subsection{Restating the within/between-design decomposition}

\citet{cadahiadelgado2026acrossdesign} defines a design $D$ as the realized configuration of price movements and other non-shock features of a panel, a shock $u$ as the residual demand disturbance, and $\theta(D)$ as the design-specific ground truth. For an estimator $\hat\theta(D,u)$, the estimation error is $e(D,u)=\hat\theta(D,u)-\theta(D)$, with design-specific conditional mean $b(D)=\mathbb{E}_u[e(D,u)\mid D]$. The within-design variance $\sigma_w^2=\mathbb{E}_D[\Var_u(e(D,u)\mid D)]$ and the between-design centring variance $\sigma_b^2=\Var_D(b(D))$ satisfy $\Var_{D,u}(e(D,u))=\sigma_w^2+\sigma_b^2$, and that analysis's central empirical finding is that $\sigma_b^2$ dominates this sum in short pricing panels. Everything in this paper concerns $\sigma_b^2$ and its multi-unit generalization; $\sigma_w^2$, the ordinary sampling uncertainty conditional on one realized design, is what conventional resampling already targets.

\subsection{From one design to many: the covariance matrix}\label{subsec:covmatrix}

Let $j=1,\ldots,N$ index \emph{candidate design-perturbation units}: in the motivating application, regions randomized under a controlled pricing experiment, though the construct generalizes to product presentations or region$\times$presentation cross-sections. Each unit $j$ carries a design-specific centring error $b_j$, with $\mathbb{E}[b_j]=\bar b_j$ (a possibly nonzero common bias, held fixed throughout) and
\[
\Sigma_{jl} \;=\; \Cov(b_j,b_l), \qquad \Sigma_{jj}=\sigma_{b,j}^2,
\]
collected into an $N\times N$ positive semi-definite \emph{design-error covariance matrix} $\Sigma$. Writing $\Sigma=\operatorname{diag}(\sigma_b)\,R\,\operatorname{diag}(\sigma_b)$ for a correlation matrix $R$ makes explicit that $\Sigma$ carries two kinds of information a passive panel bundles together and an active design can pull apart: how noisy each unit's own design is ($\sigma_{b,j}^2$), and how much any two units' designs move together ($R_{jl}$). The across-design analysis estimates exactly $R$, not just its diagonal, when it reports an intra-zone correlation of $0.603$, an inter-zone correlation of $0.572$, and a cross-UPC correlation of $0.011$ for the Dominick's Soft Drinks category it studies: the argument of this paper is that this matrix, once estimated, is a design input and not only a diagnostic.

\begin{proposition}[Portfolio form of aggregation variance]\label{prop:portfolio}
Let $w\in\mathbb{R}^N$ satisfy $\mathbf{1}'w=1$, and let $b_w=w'b$ denote the design-specific centring error of the weighted aggregate $\hat\theta_w=\sum_j w_j\hat\theta_j$, where $b=(b_1,\ldots,b_N)'$. Then
\begin{equation}
\Var(b_w) \;=\; w'\Sigma w.
\label{eq:portfolio}
\end{equation}
\end{proposition}

\begin{proof}
Immediate from the bilinearity of covariance: $\Var(w'b)=w'\Cov(b)w=w'\Sigma w$.
\end{proof}

Proposition~\ref{prop:portfolio} is the matrix form of the aggregation identity \citet{cadahiadelgado2026acrossdesign} proves for equal weights and a single equicorrelation parameter, $\Var(\bar b_k)=\sigma_b^2\frac{1+(k-1)\rho}{k}$: setting $w=\mathbf{1}/k$ and $\Sigma=\sigma_b^2[(1-\rho)I+\rho\mathbf{1}\mathbf{1}']$ in \eqref{eq:portfolio} recovers it exactly. Equation~\eqref{eq:portfolio} is also, up to notation, the variance of a portfolio's return \citep{markowitz1952portfolio}; Section~\ref{sec:theory} develops what follows from taking that correspondence seriously as an optimization problem.

\section{Theory: Combining, Purchasing, and Allocating Independent Variation}\label{sec:theory}

\subsection{Minimum-variance combination, and when the exchangeable estimator is inefficient}\label{subsec:minvar}

Given $\Sigma$, how should $N$ correlated unit-level estimates be combined into one published number?

\begin{proposition}[Minimum-variance combination]\label{prop:minvar}
Suppose $\Sigma\succ0$. The weight vector solving $\min_w w'\Sigma w$ subject to $\mathbf{1}'w=1$ is
\begin{equation}
w^\star \;=\; \frac{\Sigma^{-1}\mathbf{1}}{\mathbf{1}'\Sigma^{-1}\mathbf{1}}, \qquad \text{achieving} \qquad \Var^\star \;=\; \frac{1}{\mathbf{1}'\Sigma^{-1}\mathbf{1}}.
\label{eq:gls}
\end{equation}
\end{proposition}

\begin{proof}
See Supplement~\xref{app:proofs}.
\end{proof}

Equation~\eqref{eq:gls} is the generalized-least-squares, or minimum-variance-portfolio, rule: it is not a new statistical result, and its content here is entirely in what it is applied to. The Paule--Mandel variance-component interval of the across-design analysis implicitly uses the exchangeable working model $\Sigma=\operatorname{diag}(s_j^2)+\tau^2\mathbf{1}\mathbf{1}'$ and combines units with inverse-variance weights $w^{IV}_j\propto (s_j^2+\tau^2)^{-1}$; under exact exchangeability (equal $s_j^2$) this coincides with $w^\star$, which is why its $\sqrt{k}$ aggregation result works as reported. Once the correlation structure departs from exchangeability, as its Dominick's estimates show it does (an inter-zone correlation, $0.572$, close to the intra-zone correlation, $0.603$, is not compatible with a model in which same-zone and different-zone pairs share one common $\tau^2$ but are otherwise independent), $w^{IV}$ and $w^\star$ generally differ, and $w^{IV}$ is weakly dominated:

\begin{corollary}[Inefficiency of inverse-variance weighting under non-exchangeable correlation]\label{cor:inefficiency}
For any $\Sigma\succ0$, $w^{IV\prime}\Sigma w^{IV}\ge w^{\star\prime}\Sigma w^\star$, with equality iff $D^{-1}\mathbf{1}\propto\Sigma^{-1}\mathbf{1}$, where $D=\operatorname{diag}(\Sigma)$.
\end{corollary}

\begin{proof}
By Proposition~\ref{prop:minvar}, $w^\star$ is the unique minimizer of $w'\Sigma w$ over $\{w:\mathbf{1}'w=1\}$. The weights $w^{IV}=D^{-1}\mathbf{1}/(\mathbf{1}'D^{-1}\mathbf{1})$ satisfy the same constraint, hence attain a weakly higher value, with equality only if they coincide with $w^\star$.
\end{proof}

The loss can be large even in a stylized case: with two units sharing one design source ($\rho=0.9$) and a third independent, GLS weights cut the variance of equal weighting by $8.7\%$ without adding a single unit, because the correlated pair is partly redundant (Supplement~\xref{app:ext-threeunit}).

\subsection{Effective independent design draws}\label{subsec:neff}

\begin{definition}[Effective number of independent units]\label{def:neff}
For a homoskedastic design, $\Sigma=\sigma_b^2 R$ with $R$ a correlation matrix, define
\begin{equation}
N_{\mathrm{eff}}(R) \;:=\; \mathbf{1}'R^{-1}\mathbf{1}, \qquad \text{so that} \qquad \Var^\star \;=\; \frac{\sigma_b^2}{N_{\mathrm{eff}}(R)}.
\label{eq:neff}
\end{equation}
\end{definition}

Under equicorrelation, $R=(1-\rho)I+\rho\mathbf{1}\mathbf{1}'$, the Sherman--Morrison identity gives $N_{\mathrm{eff}}=N/(1+(N-1)\rho)$ and $w^\star=\mathbf{1}/N$ (Supplement~\xref{app:proofs}), recovering the $\sqrt{k}$ result of the across-design analysis as the case in which minimum-variance combination coincides with the simple average. $N_{\mathrm{eff}}$ is, in spirit, the effective-sample-size diagnostic of \citet{crump2009dealing} and \citet{aronowsamii2016does} that analysis invokes when it reports that Dominick's fifteen nominal price zones behave as $4.46$ effectively independent zones and its $1{,}492$ nominal UPCs as $148.4$; but it uses a spectral participation ratio, $(\sum_i\lambda_i)^2/\sum_i\lambda_i^2$ over the eigenvalues of the estimated correlation matrix, which is a different quantity from $\mathbf{1}'R^{-1}\mathbf{1}$: even under exact equicorrelation it equals $N/(1+(N-1)\rho^2)$ rather than $N/(1+(N-1)\rho)$, so the two coincide only at $\rho=0$. Re-estimating $N_{\mathrm{eff}}$ from eighteen real historical windows (Section~\ref{subsec:illustration-task4}) gives $N_{\mathrm{eff}}=8.00$ on the eight zones with usable data rather than $4.46$; Supplement~\xref{app:ext-neff} discusses the reconciliation.

\subsection{Buying variance down: information additivity under randomization}\label{subsec:additivity}

Passive units are correlated because they share calendars, cost shocks and competitive dynamics; an actively, independently randomized perturbation is constructed precisely to break that sharing. Partition candidate units into a passive set $\mathcal{J}_0$, with estimated covariance matrix $\Sigma_0$, and a set $S\subseteq\mathcal{J}_1$ selected for active perturbation at intensities $V=(V_j)_{j\in S}$, where $V_j=n_{\mathrm{moves},j}\times\mathrm{magnitude}_j^2$ is the across-design analysis's variation index, now treated as a design choice rather than an observed nuisance quantity, translating into a design variance $\sigma_{b,j}^2(V_j)=f_j(V_j)$ through a locally calibrated frontier of the form it fits, $f_j(V)=\alpha_jV^{-\beta_j}$.

\begin{assumption}[Design-induced block independence]\label{ass:blockdiag}
An independently randomized perturbation in unit $j\in S$ is generated via an exogenous treatment assignment mechanism orthogonal to the passive design process and to the randomization in every other treated unit: $\Cov(b_j,b_l)=0$ for $j\in S$, any $l\ne j$.
\end{assumption}

Assumption~\ref{ass:blockdiag} is the formal content of ``why randomize'': it is exactly what a shared promotional calendar or a common cost shock violates for passive units, and what an ex-post contamination check must protect once a perturbation is underway. Under it, $\Sigma(S,V)=\operatorname{diag}(\Sigma_0,\;\operatorname{diag}(f_j(V_j))_{j\in S})$ is block-diagonal.

\begin{proposition}[Information additivity]\label{prop:additivity}
Under Assumption~\ref{ass:blockdiag},
\begin{equation}
\frac{1}{\Var^\star(S,V)} \;=\; \iota_0 \;+\; \sum_{j\in S}\frac{1}{f_j(V_j)}, \qquad \iota_0:=\mathbf{1}_0'\Sigma_0^{-1}\mathbf{1}_0.
\label{eq:additivity}
\end{equation}
\end{proposition}

\begin{proof}
The inverse of a block-diagonal matrix is block-diagonal in the inverted blocks; apply Proposition~\ref{prop:minvar} to $\Sigma(S,V)$. Full derivation in Supplement~\xref{app:proofs}.
\end{proof}

Equation~\eqref{eq:additivity} says precision, not variance, adds: $\iota_0$, the information already resident in the passive panel, is a fixed, estimable quantity (Section~\ref{sec:sigma-estimation}), and each designed unit contributes an additional, independent term $1/f_j(V_j)$ that the committee purchases through the allocation problem of Section~\ref{sec:allocation}. This is also why the problem is tractable: absent Assumption~\ref{ass:blockdiag}, adding a unit to $S$ would require re-inverting a dense, growing matrix; under it, the passive panel's contribution is a constant the active-design problem simply adds to.

\subsection{A common-shock floor}\label{subsec:factorfloor-main}

Assumption~\ref{ass:blockdiag} asks each unit's randomization to be orthogonal not only to the passive design process but to every other treated unit's randomization. A macroeconomic shock, a simultaneous national competitor response, or a change in the manufacturer's own quota policy reaching every region in one pricing cycle violates the second requirement without necessarily violating the first. If, for $j\in S$, $b_j=\gamma_j\eta+\epsilon_j$ with one shock $\eta$ common to the cycle ($\Var(\eta)=\sigma_\eta^2$) and mutually independent idiosyncratic terms of variance $f_j(V_j)$ (Assumption~\xref{ass:factor} in the Supplement, which recovers Assumption~\ref{ass:blockdiag} at $\sigma_\eta^2=0$), a Sherman--Morrison argument gives, in the homogeneous case $\gamma_j\equiv\gamma$, $f_j(V_j)\equiv\sigma_\epsilon^2$, for the variance $\Var^\star_S$ of the minimum-variance combination of the treated units alone,
\begin{equation}
\Var^\star_S\;=\;\frac{\sigma_\epsilon^2}{|S|}+\sigma_\eta^2\gamma^2\ \xrightarrow[|S|\to\infty]{}\ \sigma_\eta^2\gamma^2
\label{eq:floor}
\end{equation}
(Proposition~\xref{prop:factorfloor}, which also gives the heterogeneous expression). Adding the passive panel's information, $1/\Var^\star(S,V)=\iota_0+1/\Var^\star_S$, so the full panel's floor is $\sigma_\eta^2\gamma^2/(1+\iota_0\sigma_\eta^2\gamma^2)$: below the treated block's own floor, but by a negligible amount whenever $\iota_0\sigma_\eta^2\gamma^2\ll1$. This is the analogue, at the level of identifying precision, of the portfolio-theory fact that diversification removes idiosyncratic but not systematic risk \citep{markowitz1952portfolio}: independently randomized units still buy precision at the rate of Proposition~\ref{prop:additivity}, but only down to a floor set by what they cannot help but share. A committee can estimate $\sigma_\eta^2\gamma^2$ from past pilot cycles and compare it with the operational target $\sigma_b\le\sigma_b^\star$ \emph{before} committing a risk budget, since, up to the passive panel's own information $\iota_0$, no amount of additional independent variation closes the gap once the shared shock alone exceeds the target. The floor bounds the consequence of such a shock, not its probability, so ex-post contamination monitoring remains necessary. A direct numerical inversion matches the closed form to within $10^{-12}$ and shows that, at $\sigma_\eta\gamma$ equal to the operational target, the treated block's precision approaches the target but never attains it at any finite number of units; with the passive panel's $\iota_0$ added, the full panel does cross it, but only at a few hundred units (Supplement~\xref{app:results-task6}).

\section{Allocating a Risk Budget}\label{sec:allocation}

\subsection{The interior, convex-cost benchmark}\label{subsec:waterfilling}

Let $\phi_j(V):=1/f_j(V)=V^{\beta_j}/\alpha_j$, which is increasing in $V$ and concave for $0<\beta_j<1$, a range that comfortably encompasses the fitted exponent ($\hat\beta\approx0.271$) of the across-design analysis. Let $c_j(V)$ denote the revenue-at-risk cost of an experimental perturbation of intensity $V$ in unit $j$, with $c_j(0)=0$.

\begin{proposition}[Water-filling allocation under convex cost]\label{prop:waterfilling}
If every $c_j$ is increasing and convex and every $\phi_j$ is concave and differentiable, the solution to $\max_{V\ge0}\sum_{j\in S}\phi_j(V_j)$ subject to $\sum_{j\in S}c_j(V_j)\le B$ satisfies, at every unit funded at a strictly positive intensity,
\begin{equation}
\frac{\phi_j'(V_j^\star)}{c_j'(V_j^\star)} \;=\; \lambda, \qquad \text{a common shadow price solving} \qquad \sum_{j\in S}c_j(V_j^\star(\lambda))=B.
\label{eq:waterfilling}
\end{equation}
\end{proposition}

\begin{proof}
Standard Lagrangian argument for a concave objective over a convex feasible region; see Supplement~\xref{app:proofs}.
\end{proof}

Proposition~\ref{prop:waterfilling} is the textbook benchmark: spend the marginal peso of risk budget wherever it buys the most marginal information, until every funded unit's bang-per-buck is equalized. It is a benchmark, not this paper's operative recommendation, because the committee-facing cost primitive developed next does not satisfy its hypothesis.

\subsection{The tolerable-volume-loss cost primitive is concave, and what that changes}\label{subsec:concave-cost}

A natural committee-facing risk quantity for a price perturbation is the tolerable volume loss $L(\Delta p)=\Delta p/(m+\Delta p)$, derived algebraically from the definition of contribution margin $m$ alone. Adopting $c_j(\Delta p_j):=\mathrm{Revenue}_j\times L_j(\Delta p_j)$ is a natural, though not the only possible, choice, since it is already the kind of number a pricing committee is asked to approve.

\begin{remark}[Concavity can change the allocation's shape]\label{rem:concave}
$L(\Delta p)$ is increasing and \emph{concave} in $\Delta p$ ($L''(\Delta p)=-2m/(m+\Delta p)^3<0$), and a single clean experimental move has $V=\Delta p^2$, so $c_j$ is concave in $V_j$ under this primitive. A sum of concave costs does not define a convex feasible region: with $c_1(V)=c_2(V)=\sqrt{V}$ and budget $B=2$, the boundary points $(V_1,V_2)=(4,0)$ and $(0,4)$ are both feasible with equality, but their midpoint $(2,2)$ has cost $2\sqrt{2}\approx2.83>2$ and is infeasible. Proposition~\ref{prop:waterfilling}'s interior characterization therefore need not locate the global optimum, and for concave-cost allocation problems optimal solutions can sit at extreme points: concentrate the budget on \emph{fewer} units pushed to a larger individual $\Delta p_j$ (bounded by an operating limit and competitive corridor, e.g.\ $\pm20\%$), rather than spreading it thinly across many units at small $\Delta p_j$ each. A non-convex feasible region does not by itself imply concentration, however. The relevant object is the composite $g_j=\phi_j\circ c_j^{-1}$, which stays concave up to the threshold of Lemma~\ref{lem:curvature}, and wherever it does water-filling remains optimal and spreading wins: for two identical units at $m_j=0.30$ and $\beta_j=0.271$, an equal split of the cost budget yields $35\%$ more information than placing it all in one unit at a budget of $5\%$ of a unit's revenue and $17.5\%$ more at $40\%$, and the gap closes (to $1.4\%$) only at $60\%$. Concentration appears in three circumstances: past the inflection point $u_j^\dagger=R_j(\tfrac12-\beta_j)$ of $g_j$, where a unit's efficient allocation is a corner (Supplement~\xref{app:ext-worstcase}); when candidates differ, since a heterogeneous pool contains units cheaper per unit of information (V4: up to $19.9\%$ more information, against under $0.5\%$ for identical units); and when a fixed per-unit cost $K_j$ makes each additional funded unit expensive (next subsection).
\end{remark}

This qualifies, rather than reverses, the naive intuition that a bigger experimentation budget should be spread over more test regions: a committee with a fixed risk appetite should prefer a small number of sharply randomized regional tests over a broad, shallow rollout when candidates are heterogeneous, when a fixed per-unit administrative cost, introduced next, is also in play, or when perturbations are pushed past the cost curve's inflection point, and not because the concave primitive alone rewards concentration. One qualification belongs here: the formal submodularity result behind greedy subset selection is proven only up to a checkable curvature threshold that the large perturbations concentration entails are the first to approach (Section~\ref{subsec:submod-general}).

\subsection{From intensity to subset selection}\label{sec:subset}

Concavity of $c_j$ alone pushes toward concentration but would still generically fund every compatible candidate at some strictly positive intensity, since $\phi_j'(0^+)=\infty$ under the power-law frontier for $\beta_j<1$. Genuine subset selection requires a fixed cost $K_j>0$ of administering any controlled test in unit $j$: coordinating with the sales team, freezing incentive discretion for the test window, monitoring for contamination. With total cost $\sum_{j}[K_j\mathbb{1}(V_j>0)+c_j(V_j)]\le B$, the problem becomes
\begin{equation}
\begin{split}
&\max_{S\subseteq\mathcal{J}_1,\ S\ \text{independent in}\ G}\ \Phi^\star(S,B), \\
&\Phi^\star(S,B):=\max\Big\{\textstyle\sum_{j\in S}\phi_j(V_j): \sum_{j\in S}c_j(V_j)\le B-\textstyle\sum_{j\in S}K_j\Big\}
\end{split}
\label{eq:subsetselection}
\end{equation}
where $G=(\mathcal{J}_1,E)$ is a \emph{spatial compatibility graph}: an edge $(j,l)\in E$ marks two candidates that cannot both be treated, because a competitor active in one geography can read and react to a price change in an adjacent one (a SUTVA-contamination concern that must be checked ex post), or because both draw on the same, singly freezable incentive pool during the test window. A feasible $S$ must be an independent set of $G$.

\begin{remark}[Submodular structure]\label{rem:submodular}
With every $K_j=0$, $\Phi^\star(S,B)$ is monotone non-decreasing in $S$, since a unit left untouched costs nothing; with fixed costs it is not, because the budget $B-\sum_{j\in S}K_j$ left for the funded units shrinks as $S$ grows, so adding a candidate can lower the value. Whether it is submodular, so that a candidate's marginal contribution shrinks as more, especially more correlated, units are already selected, is the property that, together with monotonicity and a cardinality limit, would license the greedy algorithm's $(1-1/e)$ guarantee \citep{nemhauser1978analysis}. Variance-reduction objectives of this shape are submodular in related Gaussian sensor-placement settings \citep{krausegolovin2014submodular}; Section~\ref{subsec:submod-general} replaces that analogy with proofs, isolates which of the three complicating features, the fixed cost $K_j$, the graph $G$, or the concave cost of Remark~\ref{rem:concave}, is responsible for what remains open, and shows that with fixed costs the objective is submodular but not monotone.
\end{remark}

\subsection{Submodularity and greedy selection}\label{subsec:submod-general}

The Supplement proves that $\Phi(\cdot,B)$ is monotone and submodular in the convex-cost benchmark with $K_j\equiv0$ and no compatibility graph (Proposition~\xref{prop:submod-convex}), the setting in which the $(1-1/e)$ bound of \citet{nemhauser1978analysis} applies to greedy selection under a cardinality limit on $|S|$; restoring a general graph $G$ leaves submodularity of the set function intact but replaces that limit by an independence constraint that is not a matroid, and the bound does not carry over (Corollary~\xref{cor:submod-scope}). The fixed cost $K_j$ costs more than the bound: it takes monotonicity away, because the budget $B-\sum_{j\in S}K_j$ left for the funded units shrinks as $S$ grows. With $g(u)=1-e^{-u}$, $c(V)=V$, $K_1=K_2=1$ and $B=2.5$, for instance, $\Phi^\star(\{1\},B)=0.777>\Phi^\star(\{1,2\},B)=0.442$. What survives is submodularity. The fixed cost $K_j$ and the concave cost primitive lie outside the benchmark, and both hinge on one condition on the composite $g_j:=\phi_j\circ c_j^{-1}$: the proof needs not that $c_j$ be convex, only that $g_j$ be concave.

\begin{proposition}[Submodularity with a fixed cost, under a general concavity condition]\label{prop:submod-general}
Suppose, for every $j\in\mathcal J_1$, $c_j:[0,\infty)\to[0,\bar u_j)$ (for some $\bar u_j\in(0,\infty]$) is a strictly increasing, differentiable bijection with $c_j(0)=0$, $\phi_j$ is strictly increasing, differentiable, with $\phi_j(0)=0$, and the composite $g_j:=\phi_j\circ c_j^{-1}$ is strictly concave on $[0,\bar u_j)$. Then, for every fixed total budget $B\ge0$, the value function $\Phi^\star(\cdot,B)$ of~\eqref{eq:subsetselection}, \emph{including} the fixed cost $K_j>0$ for every $j$, is submodular on the family $\{S\subseteq\mathcal J_1:\sum_{j\in S}K_j\le B\}$ of sets whose fixed costs fit the budget. It is monotone non-decreasing if every $K_j=0$, but not in general.
\end{proposition}

\begin{proof}
See Supplement~\xref{app:proofs}.
\end{proof}

\begin{lemma}[A closed-form curvature threshold for the tolerable-volume-loss primitive]\label{lem:curvature}
Let $c_j(\Delta p):=R_jL(\Delta p)$ with $L(\Delta p)=\Delta p/(m_j+\Delta p)$ (Section~\ref{subsec:concave-cost}) and $\phi_j(V)=V^{\beta_j}/\alpha_j$ with $V=\Delta p^2$ for a single clean experimental move, $0<\beta_j<1$. Then $g_j=\phi_j\circ c_j^{-1}$ is concave on $\{u:u=R_jL(\Delta p),\,0\le\Delta p\le\bar p\}$, for any operating cap $\bar p$ on $\Delta p_j$, if and only if
\begin{equation}
L(\bar p)\ \le\ \frac12-\beta_j.
\label{eq:curvature-threshold}
\end{equation}
\end{lemma}

\begin{proof}
See Supplement~\xref{app:proofs}.
\end{proof}

\begin{corollary}[What Propositions~\xref{prop:submod-convex}--\ref{prop:submod-general} jointly close, and what they leave open]\label{cor:submod-general-scope}
Under condition~\eqref{eq:curvature-threshold} at every candidate unit's own $(m_j,\beta_j)$, problem~\eqref{eq:subsetselection}, with the recommended concave cost primitive, the fixed cost $K_j>0$ \emph{and} a general compatibility graph $G$, is a submodular, but not monotone, maximization over the independent sets of $G$ whose fixed costs fit the budget, for every perturbation inside the sub-corridor $\{\Delta p:L(\Delta p)\le\tfrac12-\beta_j\}\subseteq[0,\bar p]$. It does not inherit the $(1-1/e)$ bound of \citet{nemhauser1978analysis}, which presupposes monotonicity and a cardinality constraint (an independence constraint on $G$ is neither), so greedy's performance is an empirical matter in this paper, not a guaranteed one. At $\hat\beta\approx0.271$, condition~\eqref{eq:curvature-threshold} requires $L(\Delta p)\lesssim0.229$, a strictly smaller cap than the nominal $\pm20\%$ operating limit at realistic margins (for $m_j=0.30$, it binds at $\Delta p_j\approx8.9\%$), and it binds earliest exactly where the concentration recommendation of Remark~\ref{rem:concave} pushes hardest.
\end{corollary}

This relocates the conjecture of Remark~\ref{rem:submodular} from an undifferentiated gap to a single checkable inequality per unit, strongest where the concentration recommendation is weakest.

\paragraph{Beyond the threshold.} Within the open sub-corridor $\{\Delta p_j:L(\Delta p_j)>\tfrac12-\beta_j\}$ the failure of concavity worsens monotonically with $\Delta p_j$, and Proposition~\xref{prop:submod-worstcase} shows that the worst case is attained at the corridor's upper endpoint: once a unit's efficient allocation crosses the inflection point $u_j^\dagger:=R_j(\tfrac12-\beta_j)$, the optimum of its own sub-problem is a corner at $u_j^\dagger$ or $\bar u_j$, so the decision in that regime is naturally binary. At the calibrated $\hat\beta\approx0.271$ and the $\pm20\%$ cap, every unit's corridor reaches past $u_j^\dagger$ for any margin under roughly $67\%$, so the open regime is the rule rather than a corner case. Task~7 therefore checks greedy inside it, on synthetic instances built so that every unit's corridor does so by construction: across $54$ instances, greedy averages $91.7\%$ of the exact joint optimum (range $57.7$--$100\%$), with one instance below the $1-1/e\approx63.2\%$ reference level (not a guarantee here). In $92.6\%$ of funded-unit decisions in the exact joint optimum the chosen intensity is a corridor corner or lies in the concave region; in the remaining $7.4\%$ an interior point of the open sub-corridor is jointly optimal, because a shared budget need not be supported by a single shadow price once any unit's value function is non-concave (Remark~\xref{rem:duality-gap}). This narrows rather than reopens the conjecture, which concerns the value function's submodularity rather than any one allocation, though a practitioner using shared-shadow-price water-filling can occasionally leave a small amount of value on the table here (Supplement~\xref{app:results-task8}).

\begin{figure}[tb]
\centering
\includegraphics[width=0.82\linewidth]{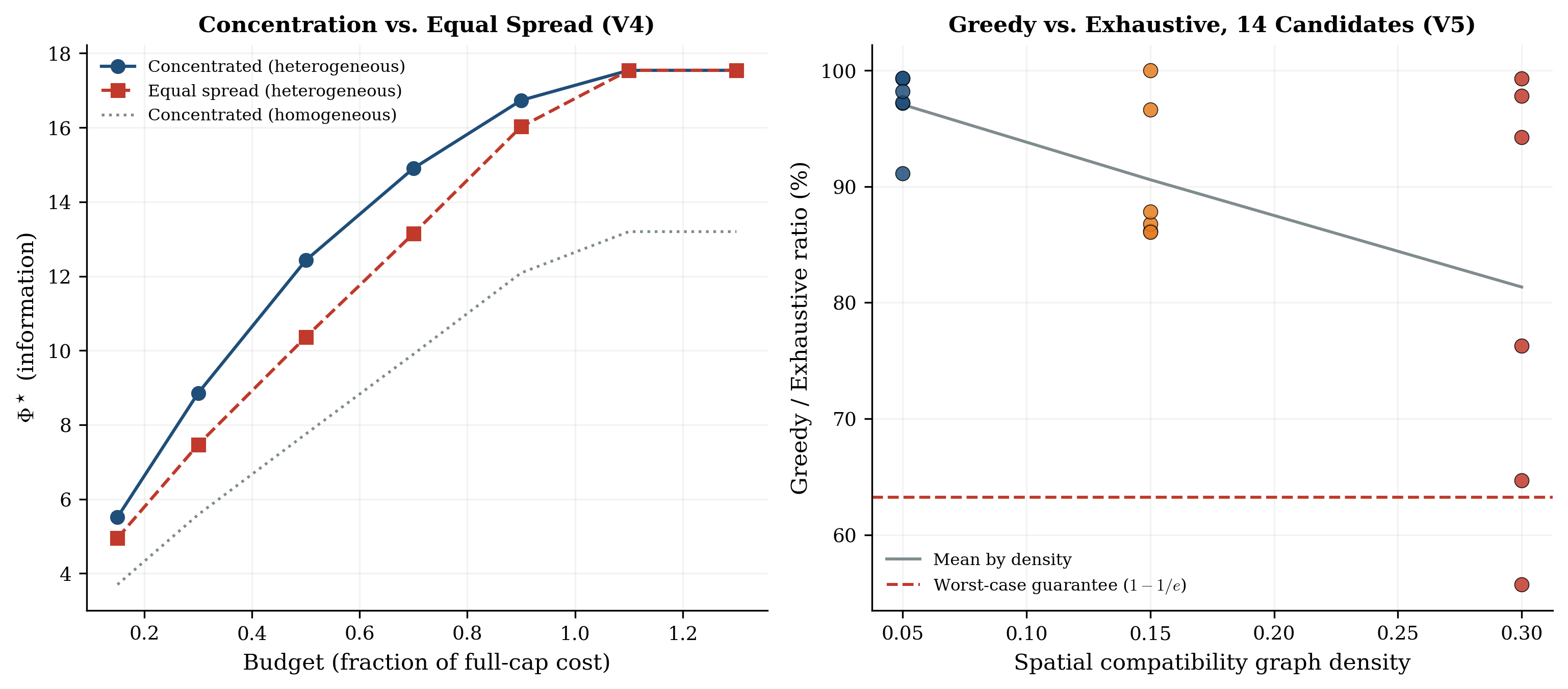}
\caption{(a)~Information $\Phi^\star$ obtained by concentrating versus spreading the budget equally across heterogeneous candidates (solid and dashed), and by concentrating it across a homogeneous pool (dotted), as a function of the budget (V4). (b)~Greedy-to-exhaustive ratio with $14$ candidates, six instances per compatibility-graph density; the solid line is the mean per density and the dashed line the $1-1/e$ reference level (V5). Synthetic panel generators; protocol and full results in Supplement~\xref{app:results}.}
\label{fig:main-allocation}
\end{figure}

\paragraph{Numerical evidence.} Figure~\ref{fig:main-allocation} shows two of the validation checks of Table~\ref{tab:summary}. With heterogeneous candidates, concentrating a fixed budget in fewer units yields up to $19.9\%$ more information than spreading it equally, while for homogeneous pools the gain is below $0.5\%$ (V4; panel~a plots the heterogeneous pool under both policies and, dotted, the homogeneous pool under concentration). Greedy subset selection averages $89.7\%$ of the exhaustive optimum at $14$ candidates, and its ratio falls as the compatibility graph densifies, with one of $18$ instances below the $1-1/e\approx63.2\%$ reference level, the bound a monotone submodular objective would carry (V5; panel~b).

\section{Guided Iteration: A Methodology for Exiting the Experimentation Zone}\label{sec:sequential}

A presentation sits in the \emph{experimentation zone} when two conditions hold: it is eligible for active perturbation (Definition~\ref{def:geographic_admissibility}), and its cumulative information $\iota^{(c)}$ falls short of an operational target $\iota^\star:=1/\sigma_{\mathrm{target}}^2$, the information at which the dispersion of the identification error reaches its target value $\sigma_{\mathrm{target}}$ (in Section~\ref{sec:illustration}, $\sigma_{\mathrm{target}}=0.088$ and $\iota^\star\approx129.1$). Sections~\ref{sec:theory}--\ref{sec:allocation} analyze the single-stage problem: given $\iota$ short of $\iota^\star$, which units to perturb, by how much, at what cost. What turns that solution into a way of exiting the zone is a sequence of them, each a deliberately designed source of independent identifying variation, chained cycle after cycle until $\iota^{(c)}$ reaches $\iota^\star$ and the panel is no longer short in the sense that matters.

We take the decision cycle to be monthly and discount the design's cumulative information at rate $\delta=0.90$ per cycle, an information weight that halves in roughly seven cycles: recent evidence should count more than evidence from two years ago, since an intervening structural break can invalidate old evidence outright.

\begin{proposition}[Discounted cumulative information]\label{prop:sequential}
Let $\iota^{(c)}$ denote cumulative information after cycle $c$, updated as
\begin{equation}
\iota^{(c)} \;=\; \delta\,\iota^{(c-1)} \;+\; \Delta\iota^{(c)}, \qquad \Delta\iota^{(c)} := \sum_{j\in S^{(c)}}\phi_j\big(V_j^{(c)}\big),
\label{eq:seqrecursion}
\end{equation}
where $\Delta\iota^{(c)}$ is the information bought by cycle $c$'s active units and the recursion starts from the passive panel's own information, $\iota^{(0)}=\iota_0=\mathbf1'\Sigma_0^{-1}\mathbf1$. With a constant per-cycle increment $\Delta\iota^{(c)}\equiv\bar\phi$, $\iota^{(c)}=\delta^{c}\iota_0+\bar\phi\,\dfrac{1-\delta^{c}}{1-\delta}$, with steady state $\iota^{(\infty)}=\bar\phi/(1-\delta)$, and $\iota^{(\infty)}-\iota^{(c)}=\delta^{c}\big(\iota^{(\infty)}-\iota_0\big)$.
\end{proposition}

\begin{proof}
Geometric series; Supplement~\xref{app:proofs}.
\end{proof}

\begin{corollary}[Cycles to the operational target]\label{cor:cycles}
If $\iota_0<\iota^\star<\iota^{(\infty)}$, the number of cycles needed to reach the operational target is $c^\star=\lceil \ln(\epsilon)/\ln(\delta)\rceil$, where $\epsilon:=\big(\iota^{(\infty)}-\iota^\star\big)/\big(\iota^{(\infty)}-\iota_0\big)\in(0,1)$ is the fraction of the initial gap to the steady state still open when the target is crossed (Proposition~\ref{prop:sequential} gives $\iota^{(\infty)}-\iota^{(c)}=\delta^{c}(\iota^{(\infty)}-\iota_0)$, so the target is reached as soon as $\delta^{c}\le\epsilon$). If $\iota^\star\ge\iota^{(\infty)}$ it is never reached, so $\bar\phi>(1-\delta)\,\iota^\star$ is necessary. Because $\epsilon$ is determined by $\iota^\star$, $\iota_0$ and $\bar\phi$, not chosen, $c^\star$ depends on all three. For a target set as a fraction $1-\epsilon$ of steady-state precision and a start at $\iota_0=0$, $c^\star$ depends on $\epsilon$ and $\delta$ alone: at $\delta=0.90$ and $\epsilon=0.10$, $c^\star=22$ cycles ($21.85$ before rounding), consistent with the roughly seven-cycle half-life this discount already implies.
\end{corollary}

\paragraph{Why the passive panel enters once.} The recursion charges the passive panel's information to the initial state, $\iota^{(0)}=\iota_0$, and lets only the active units' information into the per-cycle increment. The alternative, replenishing $\iota_0$ every cycle ($\iota^{(0)}=0$ and increment $\bar\phi+\iota_0$), is the less defensible reading, for three reasons. First, the diagnosis the paper starts from is that passive variation does not supply independent information: the passive panel's estimation error is dominated by between-design (trajectory) dispersion ($99.3\%$ of the total in the Soft Drinks re-estimation of Section~\ref{subsec:illustration-task4}), which more passive weeks do not average away, so crediting a fresh $\iota_0$ each cycle would count the same correlated price history again and again. Second, entering once keeps the steady state $\iota^{(\infty)}=\bar\phi/(1-\delta)$ a function of purchased, active information alone, so whether the target is reachable depends only on what the experiments buy, the statement the methodology is built to support. Third, it is the conservative choice: in the illustration of Section~\ref{subsec:illustration-task4} it gives $c^\star=10.15$ (cycle $11$) against $9.02$ (cycle $10$) under replenishment, a difference of about one cycle, with the passive panel contributing only $7\%$ of the replenished asymptote.

\paragraph{Iterations remaining, and what each one is for.} Corollary~\ref{cor:cycles} states $c^\star$ as a fixed count from the starting state under a constant per-cycle increment; read at any cycle $c$ already underway, $\max(c^\star-c,0)$ is the number of \emph{iterations remaining} before the zone's exit, given the discount $\delta$, the target $\iota^\star$ and the increment $\bar\phi$ (re-solving the corollary from the current state $\iota^{(c)}$ returns the same count). This count is not incidental to each cycle's design; it is what the design is chosen to move: Section~\ref{sec:allocation} selects the units and intensities for cycle $c$ so that $\Delta\iota^{(c)}$ is as large as the cycle's risk budget allows, making each cycle a substantive experiment rather than a step toward eventually accumulating enough evidence.

\paragraph{An adaptive count in place of a calendar.} A committee's default expectation of an experimentation program is a calendar: a start date, an end date, a review meeting on the books before the first test region is chosen. Guided iteration does not supply one, for a structural reason: $\max(c^\star-c,0)$ depends on $\bar\phi$, the information each cycle actually provides, both through $\epsilon$ and through the steady state $\iota^{(\infty)}=\bar\phi/(1-\delta)$ (a target above it is never reached), and $\bar\phi$ in turn depends on unit eligibility (Definition~\ref{def:geographic_admissibility}), candidate-pool heterogeneity (Remark~\ref{rem:concave}) and the approved risk budget, none known before the preceding cycles complete. The count instead updates as evidence accumulates, falling faster after a cycle that buys more information than expected and slower after one where eligibility restricted the feasible set; a calendar, by contrast, promises a date it cannot back with evidence about how much identifying variation that date's testing will have purchased (Supplement~\xref{app:ext-calendar}).

\begin{definition}[Eligibility for active perturbation]\label{def:geographic_admissibility}
Unit $j$ is eligible for active perturbation, $g_j=1$, if and only if: (i) incentive discretion in $j$ can be frozen or centralized for the duration of the test window; (ii) $j$ shares no active edge in the spatial compatibility graph $G$ (Section~\ref{sec:subset}) with any currently or concurrently treated unit; and (iii) $j$ is not currently inside a recently detected structural break in its own price-elasticity relationship.
\end{definition}

\medskip
Algorithm~1 states the resulting procedure, a transductive analogue of best-arm identification and Bayesian sequential design \citep{soare2014bestarm,chaloner1995bayesian} specialized to arms entangled by $\Sigma_0$ rather than independent.

\medskip
\noindent\textbf{Algorithm 1 (Guided iteration for price experimentation).}
\begin{enumerate}
\item At the start of cycle $c$, update $\hat\Sigma_0^{(c)}$ from all data realized so far (Section~\ref{sec:sigma-estimation}) and re-estimate $\hat\iota_0^{(c)}$, which anchors the estimate of the current state through $\delta^{c}\iota_0$.
\item Given the cycle's risk budget $B^{(c)}$ and the eligibility vector $(g_j)_{j\in\mathcal J_1}$ (Definition~\ref{def:geographic_admissibility}), solve~\eqref{eq:subsetselection} for $(S^{(c)},V^{(c)})$.
\item Execute the design; feed the realized outcomes back into the estimation of $\hat\Sigma_0^{(c+1)}$ and into each treated unit's own frontier recalibration (Supplement~\xref{subsec:performative-frontier}).
\item Update cumulative information via~\eqref{eq:seqrecursion}, report the updated iterations-remaining count $\max(c^\star-c,0)$, and advance to cycle $c+1$; stop once $\iota^{(c)}\ge\iota^\star$, marking this presentation's exit from the experimentation zone.
\end{enumerate}

\subsection{Stability of the cycle-to-cycle recursion}\label{subsec:performative}

Algorithm~1 is, in the language of \citet{perdomo2020performative}, a repeated decision rule against a distribution its own past decisions helped generate: $\hat\Sigma_0^{(c+1)}$ is estimated from a panel that includes the units and intensities chosen in cycle $c$, which were themselves chosen as a function of $\hat\Sigma_0^{(c)}$. Write the recursion as $\theta^{(c+1)}=F(\theta^{(c)})$ on the state $\theta=(\hat\Sigma_0,\{\hat\alpha_j,\hat\beta_j\}_{j\in\mathcal J_1})$. If the re-estimation map is $L_{\mathcal T,\theta}$-Lipschitz in the previous state and $L_{\mathcal T,d}$-Lipschitz in the decision, and the decision rule is $L_{\mathcal R}$-Lipschitz, then
\begin{equation}
L_{\mathcal T,\theta}+L_{\mathcal T,d}\,L_{\mathcal R}\ <\ 1
\label{eq:contraction}
\end{equation}
makes $F$ a contraction, with a unique performatively stable state reached geometrically from any starting point (Proposition~\xref{prop:performative-stability}). The condition holds for free along the passive-panel channel, because the allocation objective does not involve $\iota_0$ (Corollary~\xref{cor:performative-half}), and, along the frontier-calibration channel, the feedback from a cycle's own decision into the next calibration fades at rate $O(1/n_j)$ in the unit's accumulated calibration sample, with a constant set by the excitation of its design (Proposition~\xref{prop:performative-frontier} and Corollary~\xref{cor:performative-frontier-rate}). The calibration recursion itself is non-expansive rather than strictly contracting, so what is shown there is that the performative part of the condition vanishes, not that condition~\eqref{eq:contraction} holds literally in one step. A non-asymptotic exposure remains in the earliest cycles after a unit is first treated, while its calibration sample is still small.

\subsection{Uncertainty around ``iterations remaining''}\label{subsec:cycles-uncertainty}

Corollary~\ref{cor:cycles}'s count is exact given $\bar\phi$, $\delta$ and $\iota^\star$, but reading off where a presentation currently stands needs two estimated inputs: the passive panel's information $\hat\iota_0=\mathbf1'\hat\Sigma_0^{-1}\mathbf1$ and the per-cycle increment $\hat{\bar\phi}=\sum_{j\in S}\hat\phi_j$. The first-order variance of $\hat\iota_0$ is $2\iota_0^2/n$ for a sample covariance matrix from $n\gg N$ i.i.d.\ Gaussian observations, regardless of the correlation structure (Lemma~\xref{lem:iota-variance}; when $n$ is close to $N$ the exact variance is much larger, so the interval below is then a lower bound on the uncertainty), and each treated unit contributes an estimated $\hat\phi_j(V_j)=V_j^{\hat\beta_j}e^{-\hat a_j}$ through its recalibrated frontier coefficients, with first-order variance $\phi_j(V_j)^2z_j'\Omega_jz_j$ (Lemma~\xref{lem:phi-variance}), where $z_j:=(-1,\log V_j)'$ and $\Omega_j$ is the asymptotic covariance of $\hat\gamma_j=(\hat a_j,\hat\beta_j)'$, $\hat a_j=\log\hat\alpha_j$.

\begin{proposition}[Delta-method interval for iterations remaining, with joint frontier uncertainty]\label{prop:cycles-interval}
Under the constant-increment model of Proposition~\ref{prop:sequential} and for a target $\iota^\star$, let $r(\iota,\bar\phi):=\ln\big[(\iota^{(\infty)}-\iota^\star)/(\iota^{(\infty)}-\iota)\big]/\ln\delta$, with $\iota^{(\infty)}=\bar\phi/(1-\delta)$, be the continuously relaxed number of iterations remaining from a state $\iota<\iota^\star<\iota^{(\infty)}$; at $\iota=\iota_0$ it is the unrounded $c^\star$ of Corollary~\ref{cor:cycles}. Let $\hat\iota$ and $\hat\phi_j$, $j\in S$, be mutually independent estimators, a natural consequence of their being fit from disjoint realized histories, of the current state and of each treated unit's contribution to the increment $\hat{\bar\phi}=\sum_{j\in S}\hat\phi_j$, and write $v_\iota:=\Var(\hat\iota)$ and $v_\phi:=\sum_{j\in S}\Var(\hat\phi_j)$. Then, to first order, and without treating $\iota^{(\infty)}$ as known,
\begin{equation}
\begin{gathered}
\Var(\hat r)\ \approx\ \Big(\frac{\partial r}{\partial\iota}\Big)^{2}v_\iota\ +\ \Big(\frac{\partial r}{\partial\bar\phi}\Big)^{2}v_\phi,\\[2pt]
\frac{\partial r}{\partial\iota}=\frac{1}{\ln\delta\,\big(\iota^{(\infty)}-\iota\big)},\qquad
\frac{\partial r}{\partial\bar\phi}=\frac{\iota^\star-\iota}{\ln\delta\,(1-\delta)\big(\iota^{(\infty)}-\iota^\star\big)\big(\iota^{(\infty)}-\iota\big)}.
\end{gathered}
\label{eq:cycles-variance}
\end{equation}
At design time ($c=0$, state $\iota_0$),
\begin{equation}
v_\iota\ \approx\ \frac{2\,\iota_0^{2}}{n},\qquad v_\phi\ \approx\ \sum_{j\in S}\phi_j(V_j)^2\,z_j'\Omega_jz_j,
\label{eq:iota-total-variance}
\end{equation}
giving
\begin{equation}
\mathrm{SE}(\text{iterations remaining})\ \approx\ \sqrt{\Big(\frac{\partial r}{\partial\iota}\Big)^{2}\frac{2\,\iota_0^{2}}{n}\ +\ \Big(\frac{\partial r}{\partial\bar\phi}\Big)^{2}\sum_{j\in S}\phi_j(V_j)^2\,z_j'\Omega_jz_j}\,,
\label{eq:cycles-se}
\end{equation}
and the approximate $(1-\alpha)$ interval $\hat r\pm z_{\alpha/2}\cdot\mathrm{SE}(\text{iterations remaining})$, truncated at zero. For $c\ge1$ the same expansion holds with $v_\iota$ the variance of the estimated accumulated state, which also carries the discounted estimation error of earlier cycles' increments and is not given in closed form here.
\end{proposition}

\begin{proof}
See Supplement~\xref{app:proofs}.
\end{proof}

The interval widens as $\iota^{(c)}$ approaches $\iota^{(\infty)}$ and narrows as the passive panel's effective sample size $n$ and each treated unit's calibration history $n_j$ grow. The two channels do not weigh equally: $\partial r/\partial\iota$ is the effect of one more unit of information in hand, whereas $\partial r/\partial\bar\phi$ also moves $\iota^{(\infty)}$ and with it the target's distance to the steady state, so the per-cycle increment is typically the channel that matters (in the illustration of Section~\ref{sec:illustration}, a $10\%$ error in $\bar\phi$ moves $c^\star$ by about $1.8$ cycles, a $10\%$ error in $\iota_0$ by about $0.007$). The interval propagates both error sources jointly, including through $\iota^{(\infty)}$, though it assumes away any covariance the two channels might share through a common macro shock (Section~\ref{subsec:factorfloor-main}). When a single unit dominates $S^{(c)}$, the two terms can be compared directly to see which currently dominates the interval's width (Supplement~\xref{app:ext-cycles-interval}).

\paragraph{The interval is symmetric by construction; the risk it bounds is not.} The delta method's normal approximation is symmetric around $\hat r$ by construction, an artifact of the first-order expansion, not a claim about the sampling distribution of $c^\star$, which is strongly skewed because $c^\star\to\infty$ as $\iota^{(\infty)}$ falls to $\iota^\star$. A parametric bootstrap on the illustration (Section~\ref{subsec:illustration-task4}, Figure~\ref{fig:main-cycles}b) finds the true distribution's right tail much heavier (97.5th percentile $28.0$ against the delta-method interval's $18.2$), its lower end higher ($5.7$ against $2.1$), and $6.0\%$ of draws failing to reach the target at all. The interval's upper end should therefore be read as an understatement of how long an unlucky sequence of cycles could run, and its lower end as optimistic: the practical risk in guided iteration is skewed toward waiting longer than predicted, not exiting early.

\section{Estimating the Design-Error Covariance Matrix}\label{sec:sigma-estimation}

Every result above presupposes that $\Sigma_0$ is known; in practice it must be estimated, by one of two routes at different levels of validation.

\paragraph{Direct estimation from realized co-movement.} For units with enough independent history, $\Sigma_0$ can be estimated exactly as the across-design analysis does for Dominick's: the empirical correlation of $\Delta\log P$ across candidate units over past, non-overlapping windows. This route needs no new methodology, but it does need the regularization step of Section~\ref{subsec:robust-sigma}, which is not an optional robustness check: an unregularized plug-in estimate becomes worse than ignoring correlation altogether once its relative error passes a $10$--$20\%$ threshold.

\paragraph{Factor-model extrapolation for thin-history units.} New regions, new presentations, or units observed for too short a window to estimate their own pairwise correlations have no such history. We propose modeling pairwise correlation as a function of shared, observable exposure to common drivers (overlapping promotional calendars, shared cost-shock timing or supplier, geographic adjacency, competitor overlap), fit by ridge regression of the empirical off-diagonal correlations on an intercept, indicators for shared price zones and calendars, and a normalized distance, with the fitted correlations projected onto the nearest positive semi-definite correlation matrix. This is a proposed methodology, not a validated one: Specification~V6 (Section~\ref{sec:validation}) cross-validates it on synthetic data only.

\subsection{Regularized alternatives for a fragile \texorpdfstring{$\hat\Sigma_0$}{Sigma-hat-0}}\label{subsec:robust-sigma}

Specification~V11 shows that, regardless of which route supplies the plug-in $\hat\Sigma_0$, a relative misestimation of merely $10$--$20\%$ suffices to erase the variance-optimal combination's advantage over ignoring cross-sectional correlation entirely. Minimizing variance in practice therefore requires an estimator that trades a controlled amount of bias for reduced estimation error. We evaluate three, each a direct substitute for $\hat\Sigma_0$ in Proposition~\ref{prop:minvar} and Definition~\ref{def:neff}: Ledoit--Wolf shrinkage \citep{ledoitwolf2004honey}, a convex combination $(1-\alpha)\hat\Sigma_0+\alpha F$ with a scaled-identity target $F$ and a closed-form, data-driven intensity $\alpha$ that shrinks harder exactly when the empirical estimate is least reliable; the graphical lasso \citep{friedman2008sparse}, an $\ell_1$ penalty on the precision matrix, whose sparsity is a substantive assumption at odds with a design that deliberately selects a diverse, weakly correlated subset; and the ridge-factor model above. This regularizes the \emph{passive} panel's covariance against finite-sample noise; it does not remedy Assumption~\ref{ass:blockdiag}'s own structural fragility, addressed in Section~\ref{subsec:factorfloor-main}.

\begin{figure}[tb]
\centering
\includegraphics[width=0.66\linewidth]{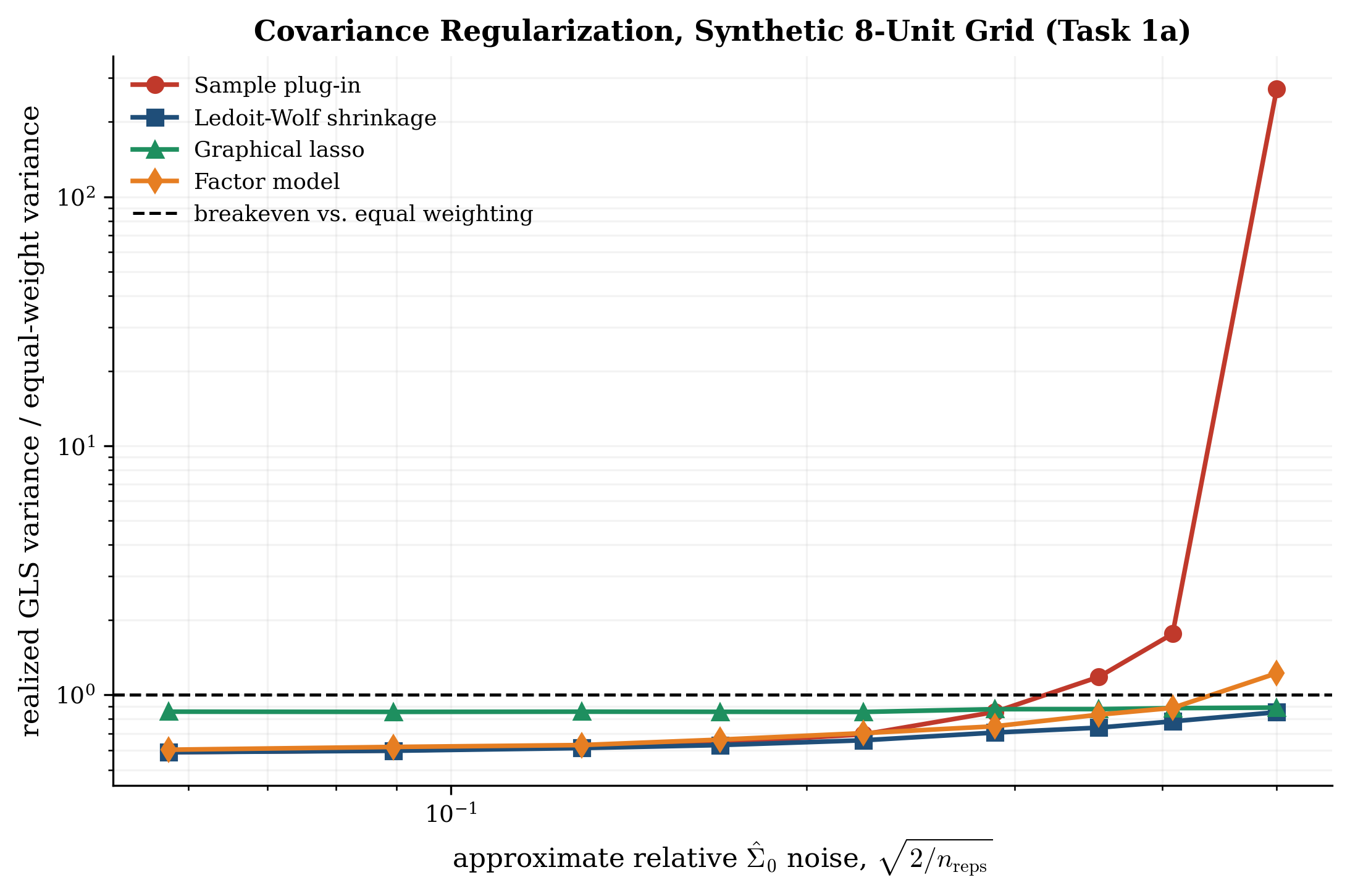}
\caption{Task~1a, synthetic eight-unit panel: realized variance of the GLS combination built from each covariance estimator, relative to equal weighting (dashed line at $1$), against the relative noise of $\hat\Sigma_0$, $\sqrt{2/n_{\mathrm{reps}}}$, with $n_{\mathrm{reps}}$ from $600$ down to $8$ (log scales). The $29\%$ and $41\%$ quoted in the text are the last grid points at which the plug-in and the factor model, respectively, still beat equal weighting.}
\label{fig:main-sigma}
\end{figure}

Task~1 reports, on a synthetic grid and on the real-calibrated setting, that Ledoit--Wolf and the graphical lasso push the plug-in estimator's collapse point furthest out: both beat equal weighting across the entire tested range, up to $50\%$ relative noise (Figure~\ref{fig:main-sigma}), against the plug-in's $29\%$, the last grid point at which it still beats equal weighting, while the factor model is last ahead at $41\%$ and behind at $50\%$. On the harder, real-calibrated eight-region case none of the three fully closes the gap to equal weighting at the noise levels tested, but Ledoit--Wolf tames the plug-in's blow-up, more than $3{,}000$ times equal weighting's variance at its worst, to within $18\%$ of it at worst. Task~5 repeats the replication-count experiment on the $15$-zone structure calibrated to the Dominick's illustration of Section~\ref{sec:illustration} (mean off-diagonal correlation $0.547$; relative noise $\sqrt{2/n}$ from $7.1\%$ to $35.4\%$): the plug-in is ahead of equal weighting up to $26.7\%$ and behind it from $30.2\%$, so its crossing lies between the two, while Ledoit--Wolf is still ahead at the largest level tried, so $35.4\%$ is a lower bound on its tolerance, not a crossing (Supplement~\xref{app:results-task1} and \xref{app:results-task6}). V11's $10$--$20\%$ belongs to the eight-region structure and a different noise model. What differs most is the headroom, GLS's variance reduction over equal weighting: $61\%$ with $\Sigma_0$ known on the $15$-zone structure, against only about $2\%$ at $10\%$ noise on the eight-region one, where a small error therefore erases it. Thresholds compare within a structure, not across structures.

\paragraph{Operationalizing structural-break eligibility (Definition~\ref{def:geographic_admissibility}(iii)).} Task~6 reconstructs the same ten-category panel at weekly, zone-level resolution with an out-of-sample fit/holdout split. $d_{\mathrm{frob}}$, the Frobenius distance between the fit and holdout sample correlation matrices, is the strongest of six candidate diagnostics screened against the cross-method instability of realized gain (the standard deviation of a category's holdout gain across the ten estimators compared; Pearson $r=0.86$, $n=10$ categories), and gating eligibility on it directly operationalizes condition~(iii). Excluding the four categories with $d_{\mathrm{frob}}>3.0$ and defaulting to equal-zone weighting there restores Ledoit--Wolf's realized gain from $-19.4\%$ blind across all ten categories to $+12.8\%$ on the six it retains, or $+7.7\%$ averaged across all ten under that deployable fallback. Nine further shrinkage-target and ensemble alternatives do not improve on Ledoit--Wolf robustly enough, across a bootstrap, a threshold sweep and an independent re-split at once, to change the recommendation: once a category's correlation structure is this unstable between fit and holdout, which shrinkage target is used is not the binding constraint (Table~\ref{tab:main-gate}; Supplement~\xref{app:results-task7}).

\begin{table}[tb]
\centering\small
\caption{Task~6 (exploratory, not specified in advance): mean realized variance reduction relative to equal-zone weighting on held-out weeks, ten Dominick's categories, weekly zone-level fit/holdout split. The gate is eligibility at $d_{\mathrm{frob}}\le3.0$. Per-category results: Supplement Table~\xref{tab:task7-diagnostics}.}\label{tab:main-gate}
\begin{tabular}{@{}lcr@{}}
\toprule
Policy & Categories scored & Mean gain (\%) \\
\midrule
Equal-zone weighting everywhere (no GLS) & $10$ & $0.0$ \\
Ledoit--Wolf GLS everywhere, blind & $10$ & $-19.4$ \\
Graphical lasso (walk-forward penalty) everywhere, blind & $10$ & $+5.6$ \\
Ledoit--Wolf where the gate passes, equal weights elsewhere & $10$ & $+7.7$ \\
\addlinespace
\emph{Memo: Ledoit--Wolf on the six eligible categories only} & $6$ & $+12.8$ \\
\bottomrule
\end{tabular}
\end{table}

\section{Illustrative Application: Dominick's Retail Data, Retrospectively}\label{sec:illustration}

The worked example below is anchored on the one category for which the across-design analysis publishes a full set of figures, Dominick's Finer Foods Soft Drinks, and reuses those figures by citation ($8{,}907{,}103$ observations, $83$ stores, $15$ price zones, $1{,}492$ UPCs, $934$ admissible UPC$\times$window strata) and applies the minimum-variance combination of Section~\ref{sec:theory} and the information-additivity result of Section~\ref{subsec:additivity} to its published correlation structure; except where Section~\ref{subsec:illustration-task4} says otherwise, the raw extract is not processed. The calculation is retrospective and counterfactual: no experiment was conducted, and the figures are hypothesis-generating rather than measurements. Section~\ref{subsec:illustration-task4} re-estimates the example on real historical windows, and Section~\ref{subsec:illustration-task5} checks across nine further categories whether the pattern is specific to this anchor category.

That analysis reports a category-level between-unit dispersion of $\hat\sigma_b=2.339$ (sieve learner; $1.215$ for the GBR learner) against an operational target of $\sigma_b\le0.088$, and, treating additional units as passively equivalent under the category's observed correlation structure, a required $707$ independent-equivalent zones. These are that analysis's published figures, cited rather than re-estimated here.

\begin{assumption}[Active randomization achieves the simulated frontier's dispersion]\label{ass:frontier-transfer}
A unit perturbed by a deliberately and independently randomized design (Assumption~\ref{ass:blockdiag}) severs the correlation responsible for real, passively observed zones' far larger dispersion, and so realizes the design variance $f_j(V_j)=\alpha_jV_j^{-\beta_j}$ that the simulated frontier predicts, rather than the passive panel's own dispersion, a factor of $13$ larger at this category's mean regime index (Remark~\ref{rem:dominicks}). This is a claim about external validity across an observational-to-experimental shift, untested here, since none of this paper's re-estimations includes a single actively randomized unit; a real pilot must confirm it before the resulting unit count is read as a budget.
\end{assumption}

\begin{remark}[What an actively designed unit might cost instead]\label{rem:dominicks}
At the category's mean regime index, $V=1.045$, the empirical frontier fit on the real extract predicts $\hat\sigma_b\approx0.180$, a prediction the real dispersion exceeds by a factor of $13$, attributable by the across-design analysis to real zones being far more correlated than the frontier's independent-design benchmark assumes. Under Assumption~\ref{ass:frontier-transfer}, combining $N$ independently designed units by simple averaging (appropriate since Assumption~\ref{ass:blockdiag} makes them mutually uncorrelated) requires $0.180/\sqrt{N}\le0.088$, i.e.\ $N\ge(0.180/0.088)^2\approx4.2$, rounding up to five units. The contrast, roughly seven hundred passively equivalent zones against roughly five actively designed ones, is offered as the order of magnitude of the value of designing variation rather than waiting for it, conditional entirely on Assumption~\ref{ass:frontier-transfer}.
\end{remark}

\subsection{A direct re-estimation on real historical windows}\label{subsec:illustration-task4}

Task~3 (Supplement~\xref{app:results-task4real}) processes eighteen non-overlapping historical windows of the anchor category's (Soft Drinks) panel from the same public archive, to estimate the design-error covariance matrix $\hat\Sigma_0$ itself rather than only its published aggregate. Of the category's fifteen nominal zones, only eight carry enough transaction data across the eighteen windows to support a zone-level elasticity estimate; the other seven are excluded rather than imputed. On these eight zones the plug-in returns $N_{\mathrm{eff}}=9.39$, again larger than the zone count, while Ledoit--Wolf saturates at $\hat\alpha=100\%$: with eighteen windows to estimate an eight-by-eight correlation matrix, the empirical off-diagonal correlations are not distinguishable from sampling noise, and shrinkage to a scaled identity gives $N_{\mathrm{eff}}=8.00$ exactly, rather than the published spectral estimate of $4.46$. This is a materially different reading, not a decimal-level reconciliation: eighteen windows are too few to resolve a correlation structure at all, and the regularizer reports that rather than fabricating one. The re-estimation nonetheless reproduces the published within/between decomposition almost exactly: a $99.3\%$ between-design share under the sieve learner against a published $99.3\%$, and $97.7\%$ under the gradient-boosting learner against $97.6\%$.

\begin{figure}[tb]
\centering
\includegraphics[width=\linewidth]{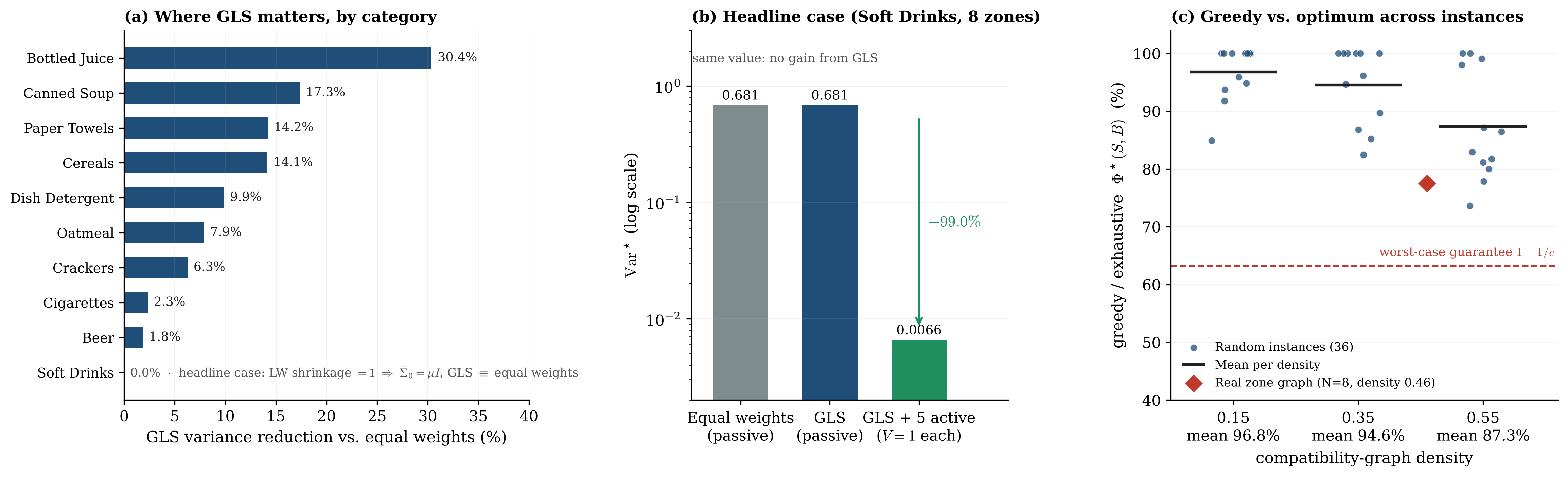}
\caption{Anchor category (Soft Drinks), real historical windows (Task~3), in the context of ten Dominick's categories. (a)~Realized variance reduction of the Ledoit--Wolf GLS combination relative to equal weights, by category (exploratory check); for Soft Drinks shrinkage saturates at $\hat\alpha=100\%$, $\hat\Sigma_0=\mu I$, and GLS coincides with equal weights ($0.0\%$). (b)~Realized $\Var^\star$ of the passive panel under equal weights and under the GLS combination ($0.681$ each), and after adding the five actively designed units of Remark~\ref{rem:dominicks} at $V=1$ each ($0.0066$, $-99.0\%$; log scale). (c)~Greedy subset selection against the exhaustive optimum, $\Phi^\star(S,B)$, on $36$ random eight-candidate instances by compatibility-graph density (Task~5; solid lines are the mean per density) and on the category's real eight-zone graph (diamond, density $0.46$, $77.5\%$); the dashed line is the $1-1/e$ reference level.}
\label{fig:main-panel}
\end{figure}

Applying Proposition~\ref{prop:minvar} to this fully shrunk $\hat\Sigma_0$ gives $\iota_0=1.468$, so the passive panel's own minimum-variance combination realizes $\Var^\star=0.681$, identical to equal weighting on the same data: a $0.0\%$ GLS gain, the arithmetic consequence of a scaled-identity $\hat\Sigma_0$ (Figure~\ref{fig:main-panel}a, bottom row). This is a boundary case of the estimation problem (Section~\ref{subsec:robust-sigma}), not a failure of Proposition~\ref{prop:minvar}: on this slice there is not enough history to certify that the zones are correlated at all. Proposition~\ref{prop:additivity} together with Remark~\ref{rem:dominicks} confirms, on the re-estimated figures, that five actively designed units still cut $\Var^\star$ from $0.681$ to $0.007$, about $99\%$ (Figure~\ref{fig:main-panel}b), since additivity does not depend on how much of $\iota_0$ reflects genuine cross-zone correlation; the $99.3\%$ quoted elsewhere is the reduction in units required, $1-5/707$. On a real eight-candidate compatibility graph built from the category's own zone adjacency (density $0.46$), greedy reaches $77.5\%$ of the exhaustive optimum at the budget implementing this design, above the $63.2\%$ reference level but in the lower part of the range for its density. Across $36$ random eight-candidate instances (Task~5; a set separate from the $N=14$ instances of Figure~\ref{fig:main-allocation}b), greedy averages $92.9\%$ of the optimum, $96.8\%$, $94.6\%$ and $87.3\%$ at densities $0.15$, $0.35$ and $0.55$, and no instance falls below the reference level (Figure~\ref{fig:main-panel}c; one real configuration, not a proof). An exploratory check, not specified in advance, applies the same real-data procedure to the nine further categories that fit (Bath Soap lacked enough usable zone-week data to support a fit): every one realizes a strictly positive GLS gain over equal weighting, from $1.8\%$ (Beer) to $30.4\%$ (Bottled Juice), mean $11.6\%$ and median $9.9\%$ across the nine (Figure~\ref{fig:main-panel}a), so Soft Drinks' $0.0\%$ reflects that category's correlation structure at this sample size, not the procedure; $c^\star$ stays within $10.00$--$10.20$ cycles in all ten categories despite a more than tenfold range in $\iota_0$, which is mechanical rather than a finding about the categories: the illustration fixes the per-cycle increment and the target across categories, only $\iota_0$ varies, and $\iota_0\ll\iota^{(\infty)}=195.8$ enters $c^\star$ only through $\ln(1-\iota_0/\iota^{(\infty)})$ (Supplement~\xref{app:results-task4real}).

\begin{figure}[tb]
\centering
\includegraphics[width=\linewidth]{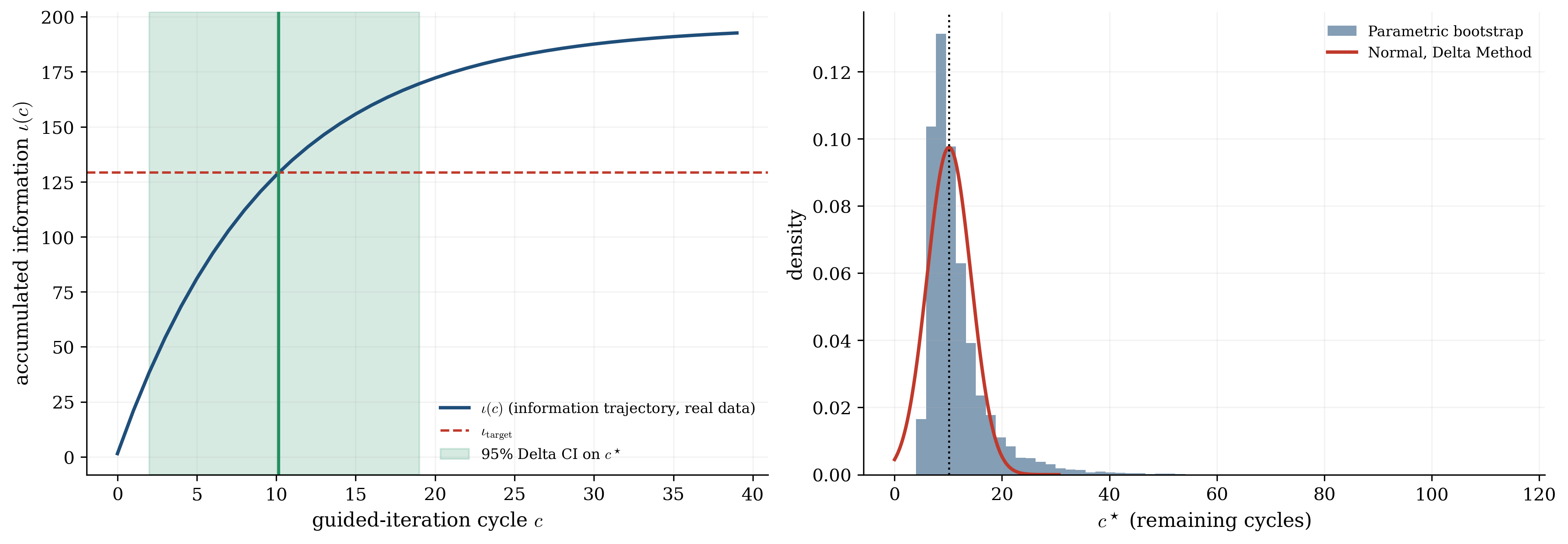}
\caption{Guided iteration on the anchor-category (Soft Drinks) illustration (Task~3). (a)~Accumulated information $\iota^{(c)}$ under the discounted recursion ($\delta=0.90$), one active unit per cycle at $V=0.45$, the operational target (dashed), $c^\star=10.15$ (vertical line) and its $95\%$ delta-method interval $[2,19]$ (shaded; it assumes a pilot of $40$ replicates). (b)~Parametric-bootstrap distribution of $c^\star$, redrawing $\hat\iota_0$ and the per-cycle increment from their asymptotic normal approximations, against the delta method's normal approximation.}
\label{fig:main-cycles}
\end{figure}

Read through Section~\ref{sec:sequential}, $\iota_0=1.468$ is also this presentation's starting state for guided iteration. Each cycle is taken to design and execute one active unit at intensity $V=0.45$, a more cautious rollout than the $V=1$ of the five-unit design above, so the per-cycle increment is $\bar\phi=1/f(0.45)=19.58$ at the calibrated frontier $f(V)=(0.182\,V^{-0.271})^2$ and the steady state is $\iota^{(\infty)}=\bar\phi/(1-\delta)=195.8$ at $\delta=0.90$, against an operational target $\iota^\star=1/0.088^2=129.1$. Corollary~\ref{cor:cycles} then gives $\epsilon=0.343$ and $c^\star=10.15$ cycles, with the target crossed at cycle $11$ once rounded up to whole pricing periods (Figure~\ref{fig:main-cycles}a). The interval of Proposition~\ref{prop:cycles-interval} is almost entirely a statement about the increment, not about the passive panel: re-estimating $\iota_0$ from eighteen windows moves $c^\star$ by a standard error of only $0.02$ cycles ($\partial c^\star/\partial\iota_0=-0.049$), whereas $\partial c^\star/\partial\bar\phi=-0.93$ per unit of information. No unit has actually been treated, so there is no frontier-calibration history to supply $v_\phi$; the illustration substitutes, by analogy with the relative error $\sqrt{2/n}$ of Lemma~\xref{lem:iota-variance}, the stylized form $2\bar\phi^2/n_{\mathrm{pilot}}$ with an \emph{assumed} pilot of $n_{\mathrm{pilot}}=40$ replicates. That gives a standard error of $4.38$ for $\hat{\bar\phi}$, of $4.09$ for $c^\star$, and a $95\%$ interval of $[2,19]$ cycles, conditional on the assumed pilot size (the interval narrows with $\sqrt{n_{\mathrm{pilot}}}$). A parametric bootstrap that redraws $\hat\iota_0$ and $\hat{\bar\phi}$ from the same asymptotic normal approximations ($n=18$ and $n_{\mathrm{pilot}}=40$) and discards the $6.0\%$ of draws for which the target is unreachable ($\iota^{(\infty)}<\iota^\star$) centers near the point estimate (median $9.9$, mean $11.5$) but is considerably more dispersed (standard deviation $6.0$ against $4.09$) and much heavier on the right, with a $95\%$ percentile interval of $[5.7,28.0]$ (Figure~\ref{fig:main-cycles}b): a reason to read a materially longer wait as more likely than a shorter one. A secondary, cost-based check, the cycle beyond which an assumed per-cycle cost equal to $3\%$ of the increment stops being repaid by marginal information, crosses only at cycle $35$; the precision criterion binds first, and cycle $11$ is the illustration's recommended stopping point.

\subsection{A real, if approximate, generalization across categories}\label{subsec:illustration-task5}

Task~4 checks whether the $707$-versus-$5$ pattern is an accident of the anchor category. Applying the same passive-to-active reduction calculation, at a fast, uniform frontier proxy rather than ten separately calibrated frontiers, to nine further Dominick's categories gives a mean reduction of $87.5\%$ (sd $17.2$pp) and a median of $97.4\%$ (IQR $23.0$pp) against the passive panel across all ten categories, with a passive-to-active equivalence multiplier of median $40\times$ (range $2$--$1{,}952\times$; the mean, $233\times$, is dominated by a handful of extreme categories): the same direction as the anchor category's own $707/5\approx141\times$, with the median about a factor of three below it, though not identical category by category (Supplement~\xref{app:results-task5}). This is a real generalization check, considerably stronger than a synthetic proxy, though still an approximation.

\section{Validation: Summary of Plan and Results}\label{sec:validation}

The validation specifies in advance of examining any data what would confirm or falsify each claim of Sections~\ref{sec:theory}--\ref{sec:sequential}; the plan of eleven pre-specified comparisons (V1--V11) is Table~\xref{tab:validation} of the Supplement. Five further evaluations extend them (Tasks~1--5), and two exploratory analyses not specified in advance (Tasks~6 and~7) are reported as such. Table~\ref{tab:summary} summarizes the sixteen comparisons; Supplement~\xref{app:protocol} gives the complete protocol and Supplement~\xref{app:results} the full results, tables and figures.

\begin{table}[htbp]
\centering\footnotesize
\caption{Summary of the sixteen validation comparisons. ``Confirmed, qualified'' denotes agreement with the theory subject to a boundary condition absent from the baseline model. Full statements, protocols and numbers: Supplement~\xref{app:protocol}--\xref{app:results} (Table~\xref{tab:verdicts}).}\label{tab:summary}
\begin{tabular}{@{}>{\raggedright\arraybackslash}p{0.07\linewidth}>{\raggedright\arraybackslash}p{0.20\linewidth}>{\raggedright\arraybackslash}p{0.48\linewidth}>{\raggedright\arraybackslash}p{0.15\linewidth}@{}}
\toprule
\# & Test & Main result & Verdict \\
\midrule
V1 & GLS vs.\ equal and inverse-variance weights & Population gain reproduces the closed form exactly; realized advantage negative until $\gtrsim150$ replications inform $\hat\Sigma$ & Confirmed, qualified \\
V2 & Information additivity & Gap $0.8\%$ under independent randomization; $17.6\%$ if units are merely \emph{labeled} independent & Confirmed, qualified \\
V3 & Optimized design vs.\ naive holdout & Beats uniform holdout by $5$--$15\%$ and passive aggregation by $21$--$37\%$ at matched budgets & Confirmed \\
V4 & Concentration vs.\ spreading & Gain up to $19.9\%$ for heterogeneous pools; $<0.5\%$ for homogeneous & Confirmed, qualified \\
V5 & Greedy vs.\ exhaustive ($N=14$) & Mean $89.7\%$ of optimum (range $55.7$--$100\%$); one of $18$ instances below $63.2\%$ & Confirmed, qualified \\
V6 & Factor-model extrapolation of $\Sigma_0$ & Cuts leave-one-store-out error by $65\%$ on a clean synthetic zone signal; real intra-/inter-zone correlations ($0.603$ vs.\ $0.572$) are far less separated & Confirmed on synthetic data \\
V7 & Sequential convergence & $90\%$ of steady state at cycle $20$ vs.\ $22$ predicted & Confirmed \\
V8 & $N_{\mathrm{eff}}$ as predictor & Predicts realized reduction within $0.5\%$; the $\sqrt{k}$ rule is off by $7.6\%$ & Confirmed \\
V9 & SUTVA-interference topology & $0$--$85.5\%$ of achievable information lost as network density grows ($85$ real stores) & Confirmed on real geography \\
V10 & Second (synthetic) category & Sieve RMSE $0.094$, correlation $0.979$, $R=20$ zones; GBR markedly less accurate & Confirmed, qualified \\
V11 & Robustness to misestimated $\hat\Sigma_0$ & Relative error of $10$--$20\%$ erases GLS's advantage over equal weights & Resolved by Task~1 \\
Task~1 & Regularized covariance estimators & Ledoit--Wolf and graphical lasso beat equal weights up to $50\%$ noise (plug-in collapses at $29\%$, factor model at $41\%$); on the real-calibrated case, tamed but not restored & Confirmed, qualified \\
Task~2 & Greedy at $N=50$--$100$ & Mean $89.9\%$ of a simulated-annealing search over $54$ configurations & Confirmed \\
Task~3 & Real windows, anchor category (Soft Drinks) & Eight of $15$ zones estimable; Ledoit--Wolf saturates, GLS gain $0.0\%$; five-unit design cuts $\Var^\star$ from $0.681$ to $0.007$ (units needed $707\to5$, $-99.3\%$); $c^\star=10.15$ cycles (one unit per cycle at $V=0.45$; exit at cycle $11$), $95\%$ CI $[2,19]$ under an assumed $40$-replicate pilot & Confirmed on real data, qualified \\
Task~4 & Ten real categories & Mean reduction $87.5\%$ (median $97.4\%$); multiplier median $40\times$ & Confirmed, qualified \\
Task~5 & Three boundary checks & Plug-in crosses equal weights between $26.7\%$ and $30.2\%$ noise, Ledoit--Wolf still ahead at $35.4\%$; floor matches closed form to $10^{-12}$; greedy $92.9\%$ of optimum on $36$ instances & Confirmed \\
\bottomrule
\end{tabular}
\end{table}

\subsection{Empirical materials}\label{subsec:materials}

Two synthetic panel generators built for this exercise (a single-region generator for the smallest closed-form checks and a thirty-region generator with heterogeneous pass-through for the combination, additivity and allocation comparisons V1--V5, V7 and V11) supply most of the simulated data; a schema-matched generator that reproduces the Dominick's database structure, with known store-to-zone assignment and true elasticities, stands in where store-level detail is needed (V6, V10). Tasks~3 and~4 process real historical windows and categories of the public Dominick's archive, and V9's compatibility graph uses that archive's store-directory file but no category price data. Every other real-data figure ($\hat\sigma_b=2.339$ and $1.215$, $R=707$, correlations of $0.603$, $0.572$ and $0.011$, and $4.46$/$148.4$ effective zones/UPCs) is cited from the across-design analysis, not re-derived. V1--V9 and V11 use one estimator throughout, a cross-fitted Ridge regression over a polynomial and Fourier sieve basis with HAC analytic variance, chosen to isolate design-variance properties; benchmarking it against a gradient-boosting learner is left to future work. Replication counts are moderate, and where results depend on them, as in V1, that dependence is reported as a finding (Supplement~\xref{app:protocol}).

\subsection{Summary of findings}\label{subsec:findings}

The theory's core predictions confirm cleanly at the population level and, for most, in realized simulation: greedy subset selection averages $89$--$93\%$ of the exhaustive or bracketed optimum across more than one hundred instances in three non-overlapping size regimes (V5, Tasks~2 and~5); the discounted recursion reaches $90\%$ of steady state two cycles ahead of schedule (V7); concentration beats even spreading once units are heterogeneous (V4); and $N_{\mathrm{eff}}$ outpredicts the naive $\sqrt{k}$ benchmark (V8). Five comparisons confirm with a substantive qualification. Additivity requires genuine, not merely labeled, independence (V2), and a factor-model covariance estimator's benefit collapses under real, closely spaced correlations (V6). Most consequentially, GLS combination's realized advantage over naive weighting is negative below roughly $150$ replications and only $10$--$20\%$ relative misestimation of $\Sigma$ erases it altogether (V1, V11); on the one real panel small enough to test directly, that advantage vanishes to exactly $0.0\%$ once eighteen windows leave Ledoit--Wolf shrinkage no choice but to saturate (Task~3). Regularization removes the plug-in's collapse within the tested range on the synthetic eight-unit panel and on the illustration's $15$-zone structure, where the plug-in collapses at $27$--$30\%$ relative noise and Ledoit--Wolf is still ahead at the largest level tried ($50\%$ and $35.4\%$ respectively); on the eight-region real-calibrated structure, where GLS's advantage is small, it only keeps GLS from becoming harmful, without eliminating the boundary condition (Tasks~1 and~5). Section~\ref{sec:limitations} treats these qualifications as open boundary conditions to be checked in practice.

\section{Limitations}\label{sec:limitations}

\paragraph{A validated theory, not yet a validated practice.} Each specification is evaluated by simulation, on synthetic data-generating processes or previously published estimates, with three exceptions: Tasks~3 and~4 process real historical windows of the Dominick's archive, and V9's compatibility graph draws on its store-directory file. Even these are retrospective and counterfactual: a real pilot, not a re-estimation on archival prices, is what would settle whether an actively randomized unit behaves as assumed (Assumption~\ref{ass:frontier-transfer}).

\paragraph{Assumption~\ref{ass:blockdiag} is an idealization.} Real randomization can leak correlation through channels this paper does not fully capture: a macro shock, a national competitor response, a policy change touching every region at once. V2's finding that the additive shortcut degrades from a $0.8\%$ to a $17.6\%$ gap once units are merely \emph{labeled} independent is a simulated instance of this failure, not a bound on how large a subtler leakage could be. The common-shock floor (Section~\ref{subsec:factorfloor-main}) turns the risk into an estimable precondition, but it bounds the idealization's consequence, not its probability; ex-post contamination monitoring remains necessary.

\paragraph{Estimating $\Sigma_0$ is bounded by category-level non-stationarity, not only by estimator choice.} The same mechanism appears at three levels of evidence. In simulation, GLS's realized advantage turns negative under limited replication and is erased by $10$--$20\%$ relative misestimation (V1, V11). On the one real panel small enough to test directly, eighteen windows leave eight of fifteen zones estimable and Ledoit--Wolf saturates, erasing the gain (Task~3); shrinking harder cannot manufacture history the panel does not have. Task~6 supplies a detector rather than a better estimator: realized Ledoit--Wolf gain is well predicted by how much a category's correlation structure shifts between fit and holdout windows, consistent with a structural break in the sense of Definition~\ref{def:geographic_admissibility}(iii). A committee should therefore treat category-level estimability, not the choice among covariance estimators, as the binding constraint.

\paragraph{The interval around the exit is partial.} Proposition~\ref{prop:cycles-interval} propagates the passive panel's and every treated unit's estimation error, including through $\iota^{(\infty)}$, but assumes away covariance the two channels might share through a common shock, and the illustration's interval rests on an assumed pilot size, not on a calibration history: no pilot exists, and the $40$ replicates are a theoretical assumption that sets the scale of the frontier channel, not an estimate. Calibrating the frontier on a real pilot and propagating its actual sampling error is the natural empirical extension, which this paper invites rather than performs. The bootstrap's heavier right tail and higher lower end mean the delta-method interval understates how long an unlucky sequence of cycles could run.

\paragraph{Submodularity is closed up to a threshold, not fully.} The proofs of Section~\ref{subsec:submod-general} cover perturbations inside a curvature threshold that binds strictly inside the nominal $\pm20\%$ cap, so the open regime is where a funded unit usually ends up. Bracketed simulation evidence over more than one hundred instances stands in for a proof there, and Task~7 targets that regime by construction (greedy at $91.7\%$ of the exact optimum). Fixed costs also remove monotonicity, so greedy's classical $(1-1/e)$ bound is not available and the case for greedy is empirical. Greedy should be read as empirically reliable throughout the operating range, and the corner-solution intuition as a good guide rather than an exact one.

\paragraph{Computational hardness and performativity.} The independent-set constraint makes exact solution of~\eqref{eq:subsetselection} NP-hard on general graphs; this paper relies on the plausibility that real compatibility graphs are structured enough (bounded degree, close to planar) for practical heuristics to perform well, and Task~2 extends the evidence to fifty to a hundred candidates (greedy at $89.9\%$ of a simulated-annealing search), which is not a proof on real graphs. Likewise, the contraction condition of Section~\ref{subsec:performative} holds for free along the passive-panel channel and closes asymptotically along the frontier-calibration channel, but a non-asymptotic exposure remains during each unit's warm-up; V7 checks the recursion's arithmetic under an assumed, not verified, stable sequence of designs.

\paragraph{Simulation scale and scope.} Replication counts are moderate; V1 shows that fine-grained quantitative comparisons, though not the qualitative conclusions, would sharpen under a larger Monte Carlo sample. The methodology and its evidence inherit the scope limits of \citet{cadahiadelgado2026acrossdesign}: stationarity up to an inflationary drift, evidence from ten product categories (Tasks~4 and~6 and the real-data check of Task~3) or, for every other comparison, from one (Soft Drinks), within a single retail chain in the 1990s, and no extension to live commercial panel data. Supplement~\xref{app:limitations} gives the extended discussion.

\section{Conclusion}\label{sec:conclusion}

\citet{cadahiadelgado2026acrossdesign} shows that short pricing panels carry little independent identifying variation and that only independently realized designs, not more rows, fix this, while leaving open how much to generate, where, and for how long. This paper responds with a guided iteration methodology: rather than a single static experiment sized once, it specifies how to iterate on prices cycle by cycle, deliberately purchasing the independent identifying variation a short, correlated panel lacks, and how to read off, from the same recursion, how many cycles remain before cumulative information reaches the target marking exit from the experimentation zone (Section~\ref{sec:sequential}), an adaptive count offered in place of a fixed calendar. It rests on supporting machinery: closed-form minimum-variance weights that generalize, and can dominate, the first analysis's variance-component estimator, and an information-additivity result that makes static allocation tractable and favors concentrated over spread designs. The secondary results of Section~\ref{sec:intro} close or narrow the gaps this machinery leaves open on its own terms.

Sixteen comparisons back these results to the extent Table~\ref{tab:summary} allows: the closed-form algebra reproduces exactly wherever it can be checked without estimation error, and the central qualitative predictions (concentration beats spreading among heterogeneous units, greedy tracks the exhaustive optimum, the sequential recursion reaches its target no later than predicted, $N_{\mathrm{eff}}$ outpredicts the naive benchmark) survive contact with simulated data. Three draw on real data: Tasks~3 and~4 process real archival windows and categories, and V9 uses the real adjacency of $85$ Chicago-area stores, locating the predicted interference cost at a plausible catchment radius of roughly $4$--$12$ miles. None of this reaches the standard a real pilot would set: every comparison used synthetic generators, published figures, or a retrospective, counterfactual re-estimation on archival prices. The roughly hundred-fold contrast of Section~\ref{sec:illustration} in the anchor category (a median of forty-fold across ten categories) illustrates both how large the prize plausibly is and how much of the argument rests on Assumption~\ref{ass:frontier-transfer}; what remains is the step retrospective evaluation cannot substitute for, a prospective, live field implementation.

\section*{Declarations}

\paragraph{Data availability.} Tasks~3 and~4 process real historical windows of the publicly available Dominick's Finer Foods archive that \citet{cadahiadelgado2026acrossdesign} also uses, and V9's compatibility graph draws on that archive's store-directory file (each store's coordinates and a relative-volume field) and no category price or transaction data. Beyond these, the paper does not access, process or redistribute the raw extract, and reports no summary statistic derived from it beyond what that analysis publishes. All remaining empirical estimates derive from the synthetic panel generators of Section~\ref{subsec:materials}, or from published benchmarks and their aligned synthetic counterparts (V6, V8, V10), as specified in Supplement~\xref{app:protocol}. Replication materials, including the data-generating specifications, the estimator definitions and the search procedures of Supplement~\xref{app:protocol-tasks}, are available from the corresponding author upon reasonable request.

\paragraph{Funding.} This research received no financial support from any public, commercial, or non-profit organization.

\paragraph{Conflicts of interest.} The author declares no competing interests.

\paragraph{Author contributions.} \textbf{Pedro Cadahia Delgado:} Conceptualization, Methodology, Formal analysis, Writing.

\paragraph{Declaration of generative AI use.} During the preparation of this work the author used Claude (Anthropic) to translate text into English and to verify the manuscript and its supplement for inconsistencies. After using this tool, the author reviewed and edited the content as needed and takes full responsibility for the content of the publication.

\bibliographystyle{plainnat}
\bibliography{referencias_designing_independent_variation}

@misc{cadahiadelgado2026acrossdesign,
  author       = {Cadahia Delgado, Pedro},
  title        = {Across-Design Uncertainty in Short Pricing Panels: Inference and Identification},
  howpublished = {arXiv preprint arXiv:2608.21334},
  year         = {2026}
}

@article{markowitz1952portfolio,
  author  = {Markowitz, Harry},
  title   = {Portfolio Selection},
  journal = {The Journal of Finance},
  volume  = {7},
  number  = {1},
  pages   = {77--91},
  year    = {1952}
}

@article{kiefer1960equivalence,
  author  = {Kiefer, J. and Wolfowitz, J.},
  title   = {The Equivalence of Two Extremum Problems},
  journal = {Canadian Journal of Mathematics},
  volume  = {12},
  pages   = {363--366},
  year    = {1960}
}

@book{pukelsheim2006optimal,
  author    = {Pukelsheim, Friedrich},
  title     = {Optimal Design of Experiments},
  publisher = {SIAM},
  series    = {Classics in Applied Mathematics},
  year      = {2006},
  note      = {Originally published by Wiley, 1993}
}

@article{chaloner1995bayesian,
  author  = {Chaloner, Kathryn and Verdinelli, Isabella},
  title   = {Bayesian Experimental Design: A Review},
  journal = {Statistical Science},
  volume  = {10},
  number  = {3},
  pages   = {273--304},
  year    = {1995}
}

@article{bojinov2021panel,
  author  = {Bojinov, Iavor and Rambachan, Ashesh and Shephard, Neil},
  title   = {Panel experiments and dynamic causal effects: A finite population perspective},
  journal = {Quantitative Economics},
  volume  = {12},
  number  = {4},
  pages   = {1171--1196},
  year    = {2021}
}

@article{athey2021matrix,
  author  = {Athey, Susan and Bayati, Mohsen and Doudchenko, Nikolay and Imbens, Guido and Khosravi, Khashayar},
  title   = {Matrix Completion Methods for Causal Panel Data Models},
  journal = {Journal of the American Statistical Association},
  volume  = {116},
  number  = {536},
  pages   = {1716--1730},
  year    = {2021}
}

@inproceedings{soare2014bestarm,
  author    = {Soare, Marta and Lazaric, Alessandro and Munos, R{\'e}mi},
  title     = {Best-Arm Identification in Linear Bandits},
  booktitle = {Advances in Neural Information Processing Systems 27 (NeurIPS)},
  year      = {2014}
}

@article{russo2018tutorial,
  author  = {Russo, Daniel and Van Roy, Benjamin and Kazerouni, Abbas and Osband, Ian and Wen, Zheng},
  title   = {A Tutorial on Thompson Sampling},
  journal = {Foundations and Trends in Machine Learning},
  volume  = {11},
  number  = {1},
  pages   = {1--96},
  year    = {2018}
}

@article{nemhauser1978analysis,
  author  = {Nemhauser, G. L. and Wolsey, L. A. and Fisher, M. L.},
  title   = {An analysis of approximations for maximizing submodular set functions {I}},
  journal = {Mathematical Programming},
  volume  = {14},
  number  = {1},
  pages   = {265--294},
  year    = {1978}
}

@incollection{krausegolovin2014submodular,
  author    = {Krause, Andreas and Golovin, Daniel},
  title     = {Submodular Function Maximization},
  booktitle = {Tractability: Practical Approaches to Hard Problems},
  publisher = {Cambridge University Press},
  year      = {2014}
}

@article{krausesinghguestrin2008sensorplacement,
  author  = {Krause, Andreas and Singh, Ajit and Guestrin, Carlos},
  title   = {Near-Optimal Sensor Placements in {G}aussian Processes: Theory, Efficient Algorithms and Empirical Studies},
  journal = {Journal of Machine Learning Research},
  volume  = {9},
  pages   = {235--284},
  year    = {2008}
}

@article{crump2009dealing,
  author  = {Crump, Richard K. and Hotz, V. Joseph and Imbens, Guido W. and Mitnik, Oscar A.},
  title   = {Dealing with limited overlap in estimation of average treatment effects},
  journal = {Biometrika},
  volume  = {96},
  number  = {1},
  pages   = {187--199},
  year    = {2009}
}

@article{aronowsamii2016does,
  author  = {Aronow, Peter M. and Samii, Cyrus},
  title   = {Does Regression Produce Representative Estimates of Causal Effects?},
  journal = {American Journal of Political Science},
  volume  = {60},
  number  = {1},
  pages   = {250--267},
  year    = {2016}
}

@article{ledoitwolf2004honey,
  author  = {Ledoit, Olivier and Wolf, Michael},
  title   = {Honey, {I} Shrunk the Sample Covariance Matrix},
  journal = {The Journal of Portfolio Management},
  volume  = {30},
  number  = {4},
  pages   = {110--119},
  year    = {2004}
}

@article{friedman2008sparse,
  author  = {Friedman, Jerome and Hastie, Trevor and Tibshirani, Robert},
  title   = {Sparse Inverse Covariance Estimation with the Graphical Lasso},
  journal = {Biostatistics},
  volume  = {9},
  number  = {3},
  pages   = {432--441},
  year    = {2008}
}

@inproceedings{perdomo2020performative,
  author    = {Perdomo, Juan and Zrnic, Tijana and Mendler-D{\"u}nner, Celestine and Hardt, Moritz},
  title     = {Performative Prediction},
  booktitle = {Proceedings of the 37th International Conference on Machine Learning (ICML)},
  pages     = {7599--7609},
  year      = {2020}
}

\end{document}